\documentclass[12pt,preprint]{aastex}
\long\def\***#1{{\scshape ***#1***}}

\makeatletter

\newenvironment{inlinefigure}{%
\def\@captype{figure}%
\noindent\begin{minipage}{0.999\linewidth}\begin{center}}
{\end{center}\end{minipage}\smallskip}
\makeatother

\lefthead{SONGAILA}
\righthead{IGM Metal Evolution}

\begin{document}

%\submitted{\today\ \  Submitted to the Astronomical Journal}
\title{The Properties of Intergalactic \ion{C}{4}  and \ion{Si}{4} Absorption, I:
Optimal Analysis of an Extremely High S/N Quasar Sample
}
\author{Antoinette Songaila\altaffilmark{1}}
\affil{Institute for Astronomy, University of Hawaii, 2680 Woodlawn Drive,
  Honolulu, HI 96822\\}

\altaffiltext{1}{Visiting astronomer, W. M. Keck
  Observatory, jointly operated by the California Institute of Technology and
  the University of California.}

\email{acowie@ifa.hawaii.edu}

\slugcomment{Submitted to Astronomical Journal; revised 7/12/05}

\begin{abstract}

We have analyzed the properties of metals in the high redshift
intergalactic medium using a novel objective pixel optical depth
technique on a sample of extremely high signal-to-noise Keck HIRES and
ESI spectra of 26 quasars between redshifts 2.1 and 6.4.  The
technique relies on using the doublet nature of the common ions
\ion{C}{4} and \ion{Si}{4} that are the principal metal tracers in the
intergalactic medium outside of the Ly$\alpha$\ forest.  Optical
depths are statistically corrected for contamination by other lines,
telluric absorption, bad pixels, continuum fitting, etc. and for
incompleteness, and we achieve in this way an increased sensitivity of
approximately 0.5 dex over previous analyses.  As with existing pixel
optical depth analyses, the method is completely objective and avoids
subjective cloud selection and Voigt profile fitting, but, unlike
existing techniques, we do not compare the ion optical depths with
\ion{H}{1} optical depths to determine the ion optical depth
distributions; we therefore avoid problems arising from different
velocity widths in the ion and \ion{H}{1}.  We have shown how the
conventional analysis can be reproduced using a percolation method to
generate pseudo-clouds from ion optical depths.  Using this set of
pseudo-clouds, we have generated \ion{C}{4} column density
distributions and have confirmed that the shape of the \ion{C}{4}
column density distribution remains essentially invariant, with slope
$-1.44$, from $z = 1.5$\ to $z = 5.5$.  This in turn confirms the lack
of redshift evolution of $\Omega({\rm CIV})$\ for $z = 2$\ to $z=5$,
both for all absorbers with column density, $\log N = 12 - 15$\ and
for stronger absorbers with $\log N = 13 - 14$.  The generation of
pseudo-clouds from the optical depth vectors also gives information on
the column density environment of a given optical depth.  We find that
for the higher resolution HIRES data there is a tight relation, $\tau
\sim N^{0.7}$, between the peak optical depth and the column density.
We have then analyzed the ion redshift evolution directly and
model-independently from the optical depth vectors themselves and show
that there is little evolution in the total amount of \ion{C}{4} from
$z = 2$\ to $z = 5$, though there is a turndown of at least a factor
of two in $\Omega({\rm CIV})$\ above $z=5$.  We do, however, see
substantial evolution in the ratio, \ion{Si}{4}/\ion{C}{4}.  In two
subsequent papers in the series, we will use this technique to
investigate what fraction of the absorbers lie in galatic wind
outflows (Paper II) and what metallicity is associated with regions of
$\tau({\rm Ly}\alpha) < 1$\ (Paper III).

\end{abstract}


\section{Introduction} \label{intro}

The availability of very efficient high-resolution spectrographs on
8--10 meter class telescopes (Vogt et al. 1994, d'Odorico et al, 2000)
provided the very high S/N observations of high redshift quasars
needed to confirm earlier suggestions (e.g. Meyer \& York 1987) that
many of the stronger lines seen in the Lyman alpha forest of quasar
spectra have associated metal absorption lines (Cowie et al. 1995,
Tytler et al. 1995). Over the past few years these metal features have
been extensively studied using Voigt profile fitting (e.g. Songaila \&
Cowie 1996; Songaila 1998; Ellison et al.\ 2000; Songaila 2001; Pichon
et al.\ 2003; Simcoe, Sargent \& Rauch 2004; Aracil et al.\ 2004) and
the so-called pixel optical depth (POD) methods in which the
metal-line optical depths are cross-correlated with neutral hydrogen
absorption (Cowie \& Songaila 1998; Songaila 1998; Aguirre, Schaye \&
Theuns 2002; Schaye et al.\ 2003; Aguirre et al.\ 2004). Down to the
sensitivity limit of these methods ($\tau({\rm Ly}\alpha) \sim 1$)
metal enrichment seems ubiquitous, with a median value of ${\rm [C/H]}
= -3.47$\ at an overdensity of $10^{0.5}$, though there is a
considerable spread of nearly a dex in the metallicity at every
overdensity and an equally significant trend with overdensity
(Shaye et al. 2003). It also appears that there is very
little change in the distribution of \ion{C}{4} absorbers over a very
wide redshift range (Songaila 2001).

The origin of these metals is still unclear: some may be in the
process of being injected into the IGM from galaxies at the redshifts
in question (we refer to this as contemporary injection) (Adelberger
et al.\ 2003, Pettini et al.\ 2003), whereas others may have been put
in place by earlier galaxy formation or by generations of population
III stars (Wasserburg \& Qian 2000; Madau, Ferrara \& Rees 2001; Qian,
Sargent \& Wasserburg 2002; Bromm, Yoshida \& Hernquist 2003;
Venkatesan \& Truran 2003; Mackey, Bromm \& Hernqist 2003; Fujita et
al.\ 2004; Daigne et al.\ 2004; Yoshida, Bromm \& Hernquist 2004).  It
is very likely that both processes contribute, and one of the primary
goals of the present series of papers is to see if we can distinguish
among systems formed by the various processes and determine if the
enrichment mechanism relates to intrinsic properties of the metal
systems such as their column density, velocity structure, or redshift.

An essential prerequisite of these analyses is, however, the existence
of extremely high signal-to-noise observations of the quasars. Over
the past years we have been obtaining very long-exposure spectra of
the brightest quasars accessible with the HIRES and ESI spectrographs
on the Keck 10-m telescope. In this set of papers we describe these
observations and then use them to reanalyse the properties of the
forest metals. In this first paper we describe a more objective
technique for analyzing the spectra which allows us to make a deeper
analysis of the data than is possible with Voigt profile fitting. In
two subsequent papers we will use this method to address two of the
outstanding problems associated with IGM metal enrichment scenarios,
namely what fraction of the absorbers lie in galatic wind outflows
(Paper II) and what metallicity is associated with regions of Lyman
alpha optical depth, $\tau({\rm Ly}\alpha) \leq 1$\ (Paper III).

Voigt profile fitting is an inherently subjective procedure, both in
the initial visual selection of the line candidates and also in the
choice of cloud model to fit the line.  The errors and incompleteness
are therefore hard to quantify. The current POD methods, while
extremely sensitive and objectively quantifiable, are sensitive only
to metal lines with other 
counterparts, either other metal lines, or,
more usually, strong hydrogen lines, and also suffer in those cases in
which the velocity structure of the metal lines differs markedly from
that of the counterpart line. Therefore we need a new technique to
analyse the absorption structure that allows us to objectively analyse
the metal data alone and to probe to the deepest possible levels that
the data allow. In this paper we develop a new method that relies
only on the doublet itself. 

The primary metal absorption features outside of the \ion{H}{1} forest
in high-redshift quasar spectra are \ion{C}{4} and \ion{Si}{4} lines
and the fact that these are doublets, and nearly all unsaturated
lines, suggests an approach to this problem.  In this paper we
describe a pixel optical depth technique in which we analyse the
spectra by looking for regions where the optical depths at the
relative wavelengths of the doublet approximate the expected 2:1
ratio; we call this the superposed pixel optical depth, or
``superPOD'' method.  We show ($\S3$) how incompleteness of the
selection arising from noise and line overlap can be quantified, and
how the completeness correction can be modelled by adding artificial
lines to the spectra. The overall analysis, which parallels the
methods used to generate number counts in images, allows us to probe a
factor of about 0.5 dex deeper in the absorption line structure than
the conventional techniques. However, even more importantly, the
method is fully automatic and therefore biases and incompleteness can
be handled properly.  We also show ($\S4$) how the conventional
analysis can be reproduced by a percolation analysis of the optical
depth results.  Finally ($\S5$) we use the data to reanalyse the
evolution of the absorption with redshift, and confirm the result that
there is little evolution of the distribution of the \ion{C}{4}
absorbers over the $z=2-5$ redshift range.  We assume throughout a
standard $\Lambda$CDM cosmology with $\Omega_m = 0.3$,
$\Omega_{\lambda} = 0.7$\ and $H_0 = 65~{\rm km\ s}^{-1}$.

\section{Observations} \label{obs}

The quasar observations are summarized in Table 1.  All of the $z<4$
quasars were observed with the HIRES spectrograph on the KeckI 10m
telescope (Vogt et al. 1994) using the red cross disperser.  Nearly
all of the observations were taken prior to the replacement of the
HIRES CCDs in 2004 but after the installation of the rotator, which
allowed the observations to be made at the paralactic angle.  A
$1.1^{\prime\prime}$ slit width was used, giving a measured resolution
of 39,000. Multiple spectrograph settings were used to give complete
wavelength coverage and observations spread over a number of years,
from the first use of HIRES in 1994 to the present, were combined to
give the final spectra.  The $z>4$ quasars were observed with the
lower resolution ESI spectrograph (Sheinis et al.\ 2000) on the KeckII
10m telescope. The slit width of $0.75^{\prime\prime}$ gives a
resolution of $\sim 5300$.

All of the reductions were carried out using an IDL software package
written by the author. More details can be found in Songaila (2001;
ESI) and Songaila (1998; HIRES). The individual spectra and
associated noise and sky files can be found at the website {\it
http://www.ifa.hawaii.edu/ $\sim$ acowie/ spectra.html}.  In each case
we have quantified the quality of the spectrum by the signal-to-noise
measured in a velocity interval of $20~{\rm km\ s}^{-1}$ at two rest
wavelengths (1250 and 1400$~\rm\AA$) in the quasar frame.  The final
signal-to-noise values measured in each of the spectra are shown in
Table 1.  The five spectra with highest quality had signal-to-noise
values near or above 200 in this interval and are marked with an H in
the table. We refer to this as the core quasar sample. These quasars
have emission redshifts from 2.72 to 3.62.

\section{Analysis Method} \label{superpod}

We restrict our analysis to absorption lines that lie more than
$1000~{\rm km\ s}^{-1}$\ longward of the quasar's ${\rm Ly}\alpha$
emission and whose redshift is more than $4000~ {\rm km\ s}^{-1}$
blueward of the quasar redshift; this avoids contamination by the
${\rm Ly}\alpha$ forest and any absorption associated with the quasar
itself. The continuum was locally fit with a 7th order polynomial,
iterating to remove absorption features, and the optical depth versus
wavelength constructed through the chosen wavelength region for each
quasar.  An upper limit of $\tau = 4.8$\ was imposed where regions
were saturated.  The optical depth vectors were smoothed  to prevent
over-sampling.  A $5~{\rm km\ s}^{-1}$\ boxcar smoothing was chosen,
which roughly matches the half widths of features seen in the metal
forest. A sample region of an optical depth vector can be found in
Figure~\ref{fig:selec}.

The distribution of optical depths in the core quasar sample is shown
in Figure~\ref{fig:distrib}.  In the selected wavelength region, 3.4\%
of the pixels in the optical depth vectors in the core sample have
optical depths above 0.1 and about a third lie above 0.005.  Much of
this structure is produced by \ion{C}{4}, but there are contributions
also from \ion{Si}{4}, \ion{N}{5} and lower ionization lines as well
as incompletely removed telluric features, continuum fitting errors,
etc.  The total optical depth can be used to provide an upper estimate
of $\Omega({\rm C~IV})$.  Features with $\tau > 0.005$ correspond to
$\Omega({\rm C~IV}) < 5.8 \times 10^{-8}$\ in the five quasars of the
core sample.  Much of this comes from higher optical depth regions;
$\tau > 0.1$\ gives a contribution to this number of $4.6 \times
10^{-8}$.  Portions of the spectrum with $0.01 < \tau < 0.1$\ give
$1.0 \times 10^{-8}$.  As we shall subsequently show, slightly more
than $50\%$\ of the total absorption comes from the \ion{C}{4} doublets.

The doublet structure of the \ion{C}{4}, \ion{Si}{4} and \ion{N}{5}
lines can now be used to reduce the contamination by other lines or
artifacts.  For each wavelength position in the optical depth vector
we can look at the corresponding wavelength position of the second
member of the doublet and see if the optical depth is in the correct
ratio of roughly 2:1.  We can then restrict the optical depth vector
to positions where the condition is satisfied and set the remaining
positions of the vector to zero. We refer to this final vector as the
cleaned optical depth.  In general the measured doublet ratio will
only approximate the true value because of noise, errors in the
continuum fit, line contamination, etc., and we need to specify a
selection parameter, that is, the range of acceptable ratios, which we
refer to as the doublet ratio window.  Too wide a choice of window
will result in overselection of false positive signal whereas too
narrow a choice will eliminate too many real systems, especially at
low optical depth.  We have chosen a doublet ratio window of 1:1 to
4:1 in the present work.  We will discuss reasons for this choice
below.

The procedure is illustrated in Figure~\ref{fig:selec}.  The bottom
panel shows the \ion{C}{4} $\lambda1548~{\rm\AA}$\ optical depth as a
function of wavelength for a $50~{\rm\AA}$\ sample of spectrum, with
one strong \ion{C}{4} line.  In the middle panel the \ion{C}{4}
$\lambda1550~{\rm\AA}$\ optical depth is shown in the rest frame of
the \ion{C}{4} $1548~\rm\AA$\ line.  Both vectors are
individually smoothed as previously described.  Finally, in the
top panel, the result of the optical depth selection criterion of
$0.25 < \tau(1548)/\tau(1550) < 1$ is shown and confirms the strong
system as a real \ion{C}{4} doublet while removing much of the weaker
structure seen in the raw \ion{C}{4} $1548~\rm\AA$\ optical depth
vector.

Figure~\ref{fig:tramon} shows the procedure in a different way. Here we have plotted
on the x-axis the \ion{C}{4} $\lambda$1548~$\rm\AA$\ optical depth
over the wavelength range, and on the y-axis the corresponding optical
depth at the second member of the CIV doublet. We label these as
$\tau(1548)$ and $\tau(1550)$ respectively.  Features that correspond
to uncontaminated CIV absorption will lie along the diagonal line,
which is indeed seen to be heavily populated.  The parallel lines show
the \ion{C}{4} doublet ratio window, here set to 0.25 -- 1.  Clearly a
narrower window could be used at high optical depths but since there
is little contamination there is no necessity for this. At lower
optical depth a narrower window would result in a larger proportion of
the points being scattered out of the window. In this diagram,
absorption features that are not \ion{C}{4} fall in the lower right
corner if they lie at the 1548~$\rm\AA$ position and in the upper left
corner if they are at the 1550~$\rm\AA$ position. It is these features
that are cleaned from the sample by the doublet ratio selection.  The
final cleaned optical depth is $\tau(1548)$ for all objects that lie
within the tramlines and zero for those outside.

A major advantage of this method is that it allows an objective
analysis of the degree of contamination since we can compare the
optical depth distribution for a true doublet --- \ion{C}{4}, say ---
with a similarly generated optical depth distribution for artificial
``doublets'' with slightly different wavelength spacing.  This is
illustrated in Figure~\ref{fig:tramoff} where we show the plot of $\tau({\rm
artificial})$ versus $\tau(1548)$ for an artificial doublet with a
12~$\rm\AA$ spacing. There are only a small number of contaminating
points within the tramlines. It is also visually obvious by comparing
Figure~\ref{fig:tramon} ( the true doublet) with Figure~\ref{fig:tramoff} (the artificial doublet)
that the signal of real doublets (the difference between the number of
true and artificial position) persists at least to $\tau({\rm
CIV~1548}) \sim 0.01$.  The aim of the method is to find a statistical
measure of the signal down to the deepest possible level.

Since the noise levels vary from spectrum to spectrum
and since telluric contamination may be a function of the wavelength
position, this type of analysis is best carried out individually on
each quasar spectrum with the individual results subsequently being
combined. Figure~\ref{fig:odonoff} shows the results
for the $z_{\rm em} = 3.03$\ quasar HS1946+7658, which has a S/N at
the low end of the high S/N quasar core sample.  The solid histogram
with $\pm 1~\sigma$\ error bars in Figure~\ref{fig:odonoff} is the
\ion{C}{4} $\lambda1548~{\rm\AA}$\ optical depth distribution for
lines in the spectrum selected as previously discussed.  The dashed
histogram is the average optical depth distribution retrieved from 20
artificial doublets with incorrect doublet separations (approximately
that of the \ion{C}{4} doublet) to model contamination of real
\ion{C}{4} doublets by interloper lines, bad pixels, continuum fitting
structure, telluric absorption, etc.  The difference of the two
histograms shown in Figure~\ref{fig:odsignal} is the \ion{C}{4}
$\lambda1548~{\rm\AA}$\ optical depth distribution statistically
corrected for contamination.  There is significant positive signal
down to at least $\tau(1548) \sim 0.005$.

The distributions must also be corrected for incompletness since lines
that are contaminated by other absorption features will be missed and
since noise and structure may scatter weak lines outside the selection
window.  That is, as well as knowing what the signal is, we also need
to have an accurate estimate of the line recovery rate.  We do this in
standard fashion by introducing artifical lines into the real spectra
and tracking the number of recoveries as a function of the \ion{C}{4}
optical depth.  The lines are added in a Monte Carlo fashion
drawn from the observed core sample \ion{C}{4} distribution.
The line widths are chosen to match the range seen in the observed
absorption features but the method is not particularly sensitive to
this choice. Figure~\ref{fig:recovery} shows a sample strip of
spectrum into which two artificial lines with $\log N({\rm CIV}) =
11.75$\ and two with $\log N({\rm CIV}) = 12.75$\ have been
introduced.  The panels have the same meaning as in
Figure~\ref{fig:selec} and the short vertical lines in the top panel
indicate the wavelengths of the introduced lines.  This is the line
strength at which recovery starts to break down, going from 100\% at
$\log N({\rm CIV}) \sim 12$ to $\sim 75\%$\ at $\log N({\rm CIV}) =
11.75$.  The recovery rate is also a function of the resolution of the
spectra and contamination is more severe and recovery poorer in the
lower resolution high-redshift ESI spectra.  The final recovery rate
versus optical depth is shown in Figure~\ref{fig:rate}.  This gives
the incompleteness correction as a function of optical depth for the
individual quasar spectrum, which can then be divided into the
observed optical depth of Figure~\ref{fig:odsignal} to obtain the true
optical depth distribution. For the high S/N core quasar sample,
incompleteness reaches 50\%\ at an optical depth of $\sim 0.017$.

\section{Voigt profile fitting emulation} \label{distrib}

The cleaned optical depth vector contains all the information present
in the traditional cloud analyses and in general it is best to use
this directly to analyse the absorption structure.  However, it is
interesting and very straightforward to relate the present type of
analysis to the analysis by Voigt profile fitting and to see how well
we reproduce the results previously obtained by this method.  In
addition, we can use the method to characterize the local environment
in which a given optical depth point lies.

We can use the superPOD method to emulate the usual Voigt profile
fitting analysis of the spectra in an objective way.  The basis of
this is to generate `pseudo-cloud complexes' by combining neighboring
optical depths with detections into a single line profile. We do this
by first finding the maximum optical depth in the true optical depth
vector. We then define the cloud complex as all positions with
non-zero optical depths that are connected to the primary position,
with gaps where the optical depth is zero being less than 100 km
s$^{-1}$.  An example of a complex picked out in this way is shown in
Figure \ref{fig:pseudoex}.  Positions already allocated to cloud
complexes are eliminated from the cleaned optical depth vector and the
process repeated. The column density is computed for each cloud
complex from the integral of the optical depths included in it, with
each optical depth being weighted by the completeness and
contamination correction for that optical depth.

The exact definition of `neighboring' is somewhat arbitrary in the use
of the $100~{\rm km\ s}^{-1}$\ window to define the cloud size.  In
general, the wider the `cloud' window, the more likely one is to lose
lower column density clouds as individual entities: they would be
incorporated into broader `cloud complexes'.  However, in practice,
for reasonable values of the window the results are not too sensitive
to this choice.  We illustrate this in Figure~\ref{fig:window}, where we show the
number of clouds found as a function of column density for velocity
windows of 75 -- 300~${\rm km\ s}^{-1}$.  The wider window reduces the
number of independent low column density complexes but the effect is
not large.

In Figure~\ref{fig:c4dist} we show as open squares the raw \ion{C}{4} column density
distribution of pseudo-clouds obtained from \ion{C}{4} optical depths
retrieved by the superPOD method from the core quasar sample at an
average redshift of 2.7.  The filled squares with $1~\sigma$ error
bars show the distribution corrected for incompleteness and bias.  For
comparison, the open circles show the results of a traditional
analysis of the same set of objects, conducted in the usual way with
visual identification of lines and subsequent manual Voigt profile
fitting.  The best-fit line has a slope of $-1.44$, similar to the
result of Ellison et al.\ (2000) at the same redshift.  This is the
dashed line in Figure~\ref{fig:c4dist}.  The solid line has a slope of $-1.7$, based
on fitting only to complexes with $N > 10^{13}~{\rm cm}^{-2}$.

Using the various quasar samples of Table~1 at low and high redshifts,
we have constructed in this way the \ion{C}{4} column density
distributions in the redshift range 1.5 -- 5.5.  In general, the
superPOD method using pseudo-cloud generation reproduces the
invariance of \ion{C}{4} column density distribution function with
redshift already determined by traditional methods (Songaila 1997,
2001; Pettini et al. 2003) and, with an extra sensitivity of about 0.5
dex over these previous results, the redshift invariance of the
distribution functions is seen very clearly.  Following the methods
used in the previous studies, we can also construct $\Omega({\rm
CIV})$\ as a function of redshift, and this is shown in Figure~\ref{fig:c4omega}.
This agrees very well with the traditional studies in showing that
$\Omega({\rm CIV})$\ remains nearly invariant between redshifts 2 and
5.  This remains true whether counting stronger absorbers (open
squares: $13 < \log N({\rm CIV}) < 14$) or all absorbers (filled
squares: $12 < \log N({\rm CIV}) < 15$).

We can extend the method by computing the column densities of lines
corresponding to other species for the range of velocities in each
\ion{C}{4} - selected cloud complex. For doublets such as \ion{Si}{4}
we use the cleaned optical depth for that species while for singlets
such as \ion{C}{2} we simply use the raw optical depth.  These ion
ratios are shown as a function of redshift for the stronger cloud
complexes in Figures~\ref{fig:c2c4} and \ref{fig:si4c4}. In both
figures, the small filled squares denote the ion ratios determined
from the \ion{C}{4} pseudo-clouds (with width of $100~{\rm km\
s}^{-1}$), whereas the open diamonds show the median ion ratios in
each redshift bin determined directly from the optical depths.
Figure~\ref{fig:c2c4} demonstrates that neither the range nor the
median value of \ion{C}{2}/\ion{C}{4} changes significantly as a
function of redshift, agreeing with Schaye et al.(2003).  However, the
same is not true of \ion{Si}{4}/\ion{C}{4}, a result which a
Kolmogorov-Smirnov test (Figure~\ref{fig:ks}) shows to be highly
significant.  The large open squares in Figure~\ref{fig:si4c4} denote
pseudo-clouds with high \ion{C}{4} column density.  The K-S test is
based on all the \ion{Si}{4} and \ion{C}{4} systems, including these
strong ones.   However, it is dominated by the larger numbers
of smaller column density systems.  On the other hand, evolution in
both $\Omega ({\rm CIV})$ and $\Omega ({\rm SiIV})$ is dominated by
the high column density end of the distribution.   We discuss
this further, using optical depth vectors, in Paper~II, and we
postpone the interpretation to that paper.

In Figure~16 we show the distribution of the optical depths as a
function of the column density of the cloud complex in which they
lie. For the HIRES data there is a tight relation ($\tau \sim
N^{0.7}$) between the peak optical depth and the column density, which
is washed out in the lower resolution ESI data.  In paper~II we shall
revisit this relation, which imples that higher column density clouds,
or, equivalently, systems with larger peak optical depths, 
are also wider in velocity space.

\section{Optical depth analysis} \label{anal}

It is encouraging that there is such good agreement between
traditional Voigt profile fitting methods and the objective, but still
slightly arbitrary, method of generating pseudo-cloud complexes.
However, the real merit of the superPOD technique lies in the fact
that the cleaned optical depth vectors contain all of the information
about the amount of material, its distribution in strength, and the
velocity structure of the absorption, and may be used directly to
characterize these without ever resorting to arbitrary groupings.

In the present paper we use the optical depth vectors to determine the
metallicity evolution directly from \ion{C}{4} and \ion{Si}{4} optical
depths, postponing a discussion of the velocity structure and the
ionization balance evolution to Paper II.  Our results are shown in
Figure~\ref{fig:odomega} for \ion{C}{4} (filled squares) and \ion{Si}{4} (filled
diamonds) in four redshift bins, [1.5,2.5], [2.5,3.5], [3.5,4.5], and
[4.5,5.5]. We see about a factor of two turndown in \ion{C}{4} at $z =
5$\ and perhaps a larger turndown in \ion{Si}{4} at the same redshift.
The S/N of the data is declining at the highest redshifts and the sky
systematics are higher but, even given these caveats, the slight
turndown at these redshifts seems real and could correspond to a
decline in the metal density or a change in the ionization balance
(Shaye et al.\ 2003). Irrespective of this point there is clearly a
significant metal density in place at $z=5$.

In Figure~18 we show the optical depth distributions of \ion{C}{4}
over the same redshift ranges.  We have divided the quasar sample into
five sets of five quasars, ordered by redshift, which are shown in the
five panels, (a) to (e).  (We omitted the quasar PSS0747+443 and
divided the remainder to give the five groups.)  The redshift range of
\ion {C}{4} absorption in each set of quasars is marked on the panel.
The three highest redshift panels are based on ESI data and the two
lowest on HIRES data.  In each case we have overplotted the
distribution obtained from the core high S/N quasar sample.  For the
ESI data we have smoothed the comparison sample to the ESI resolution.
Rather than show Poisson error bars, which underestimate the errors
because of correlation effects, we have shown in each case the range
of the five quasars.  There is very little change in the shape of the
distribution over the redshift interval, which is consistent with the
results of the Voigt profile analyses.

The slow change in the overall ion densities and the invariance of the
distribution functions appear to suggest that contemporary injection
must not be a dominant mechanism for producing the observed absorption
lines, since we would in that case expect the features to track the
evolution of the star-forming galaxies and to have some dependence on
the changing properties of these objects. Indeed, even if this were
the only mechanism present, we would expect that metals ejected by
previous generations of star-foming galaxies would build up in the
intergalactic gas, and that as time progressed the relative numbers of
these versus those being formed by contemporary injection would
change, so producing evolution in the column density distribution with
redshift. However, such effects could be concealed if the ejected
material becomes too hot to be observed as \ion{C}{4} (Aguirre et al.\
2004), and in the second paper of this series we will look at other
approaches to determining the fraction of the lines that arise in
contemporary injection.

\section{Conclusion} \label{conc}

We have described a novel objective pixel optical depth technique,
``superPOD'', that uses the doublet nature of the common ions
\ion{C}{4} and \ion{Si}{4}, that are the principal metal tracers in the
intergalactic medium outside of the Ly$\alpha$\ forest, to analyze the
properties of metals in the high redshift intergalactic medium using a
sample of extremely high signal-to-noise Keck HIRES and ESI spectra of
26 quasars between redshifts 2.1 and 6.4.  The method is completely
objective and, using statistical corrections for contamination by
other lines, telluric absorption, bad pixels, continuum fitting,
etc. and for incompleteness, gives us an increased sensitivity of
approximately 0.5 dex over previous analyses.

As a check on the method, we have reproduced the conventional analysis
using a percolation method to generate pseudo-clouds from ion optical
depths.  Using this set of pseudo-clouds we confirm that the
\ion{C}{4} column density distribution remains essentially invariant,
with slope $-1.44$, from $z = 1.5$\ to $z = 5$, which confirms
the lack of redshift evolution of $\Omega({\rm CIV})$\ for $z = 2$\ to
$z=5$\ previously determined from conventional Voigt profile fitting
analysis.

The pseudo-clouds also give information on the column density
environment of a given optical depth: for the higher resolution HIRES
data there is a tight relation, $\tau \sim N^{0.7}$, between the peak
optical depth and the column density.

Using the optical depths directly, we show that there is little
evolution in \ion{C}{4} and \ion{C}{2}/\ion{C}{4} from $z
= 2$\ to $z = 5$, though there is a turndown of at least a factor of
two in $\Omega({\rm CIV})$\ and $\Omega({\rm SiIV})$\ above $z=5$; we
do, however, see substantial evolution in \ion{Si}{4}/\ion{C}{4}.

We will use this technique in two subsequent papers to investigate
what fraction of the absorbers lie in galatic wind outflows (Paper II)
and what metallicity is associated with regions of $\tau({\rm
Ly}\alpha) < 1$\ (Paper III).

\acknowledgments 

I would like to thank X. Fan and R. Becker for providing the
coordinates of some of the $z > 6$\ quasars prior to their
publication.  This research was supported by the National Science
Foundation under grant AST 00-98480.

%  TABLE 1
\begin{deluxetable}{lccccc}
\small
\tablewidth{350pt}
\tablenum{1}
\tablecaption{The Quasar Sample \label{tbl:1}}
\tablehead{
\colhead{Quasar} & \colhead{$z_{em}$}  & \colhead{S/N $1250~{\rm\AA}$} 
& \colhead{S/N $1400~{\rm\AA}$} & \colhead{Instr.$^{(a)}$} &
\colhead{Sample$^{(b)}$} }
\startdata
 SDSS1148+52   &  6.39000  &    80   &   -5   &  E & ? \nl
 SDSS1048+46   &  6.23000   &   75   &   -5   &  E & ? \nl
 SDSS1306+03   &  5.98500   &   95   &   -5   &  E & 4 \nl
 SDSS0836+00   &  5.76500   &  130   &   55   &  E & 4 \nl
 SDSS1044-01   &  5.75500   &   70   &   25   &  E & 4 \nl
 SDSS1204-00   &  5.05500   &  105   &   80   &  E & 4 \nl
 SDSS0338+00   &  4.99000   &  115   &   45   &  E & 4 \nl
 SDSS1737+58   &  4.84000   &  100   &  115   &  E & 3,4 \nl
 SDSS2200+00   &  4.76300   &  125   &   65   &  E & 3,4 \nl
 BR1202-0725   &  4.60000   &  135   &   95   &  E & 3 \nl
 BR0334-1612   &  4.36000   &  130   &   85   &  E & 3 \nl
 BR0353-3820   &  4.55500   &  100   &   60   &  E & 3 \nl
 BR2237-0607   &  4.55000   &  230   &  220   &  E & 3 \nl
 PSS0747+443   &  4.43200   &   95   &   60   &  E & 3 \nl
 BRI0952-011   &  4.40800   &  100   &   90   &  E & 3 \nl
 PSS0926+305   &  4.19000   &  135   &  120   &  E & 3 \nl
 BR2237-0607   &  4.55000   &   55   &   35   &  HH & ? \nl
 Q1422+2309    &  3.62000   &  340   &  285   &  H & H,2 \nl
 HS0741+4741   &  3.22000   &  210   &  185   &  H & H,2 \nl
 Q0636+680     &  3.18000   &  220   &  165   &  H & H,2 \nl
 HS1946+7658   &  3.03000   &  210   &  200   &  H,HH & H,2 \nl
 HE2347-4342   &  2.88000   &  115   &  105   &  H & 1 \nl
 HS0119+1432   &  2.87000   &  115   &   95   &  H & 1 \nl
 HS1700+6416   &  2.72000   &  205   &  175   &  H & H,1 \nl
 HE1122-1648   &  2.40000   &   60   &   75   &  H & 1 \nl
 HS1626+6433   &  2.31000   &   55   &   80   &  H & 1 \nl
 Q1331+170     &  2.08000   &   30   &  110   &  H & 1 \nl

\enddata

\tablenotetext{(a)}{E : ESI spectrograph on the Keck~II telescope; H :
Hires spectrograph on the Keck~I telescope;  HH : HIRES with upgraded detector}

\tablenotetext{(b)}{H : high S/N sample;  1,2,3,4 : samples with
$\langle z\rangle = 2.2,2.8,3.9,4.5$}

\end{deluxetable}

\newpage
%
% Figure 1  Basic doublet selection
%
\begin{inlinefigure}
\includegraphics[angle=90,scale=.6]{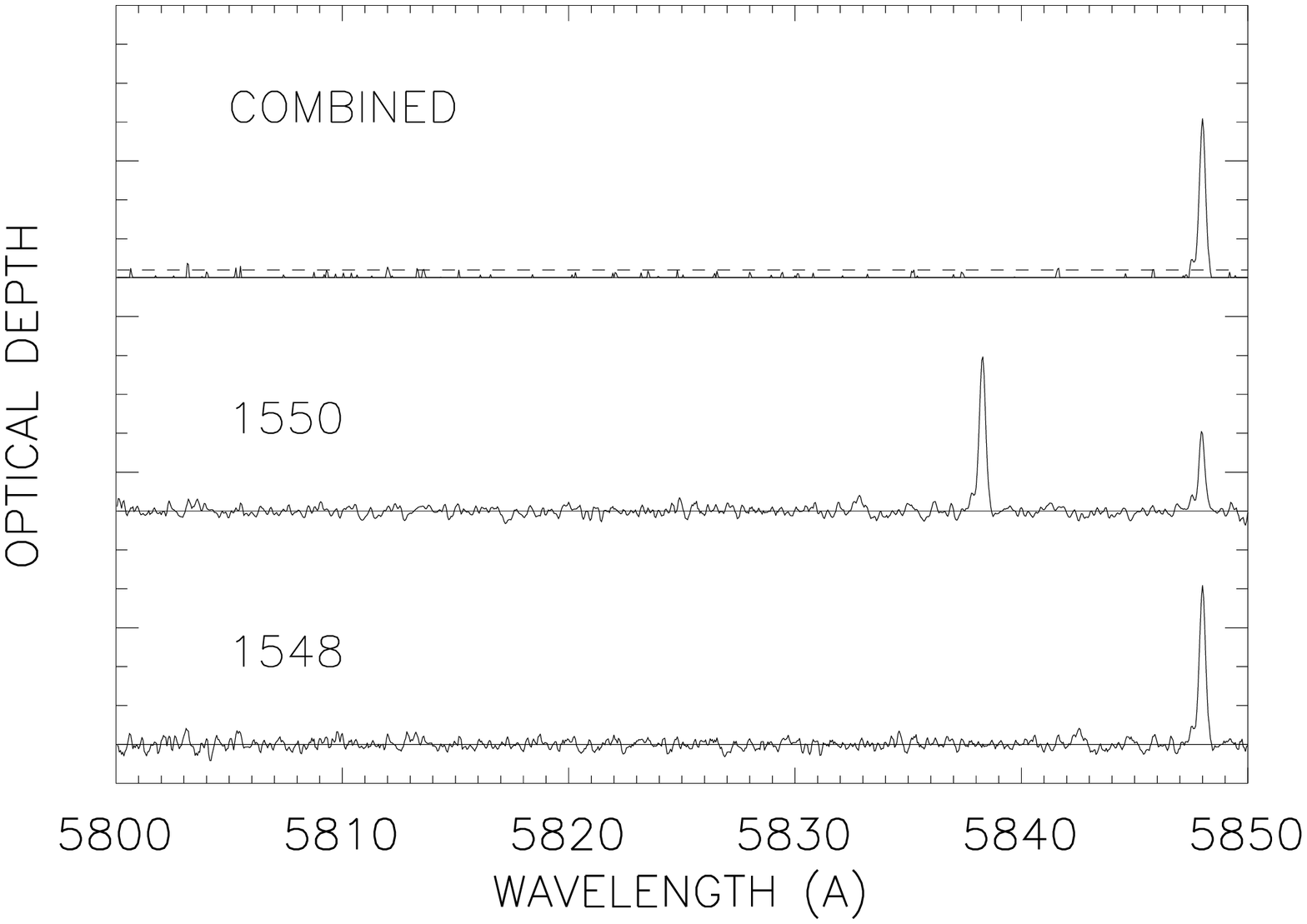}
\vspace{6pt}
\figurenum{1}
\caption{ 
Illustration of the doublet selection method based on a portion of the
spectrum of the quasar HS1946+7658.  {\it Bottom panel\/}: \ion{C}{4}
$\lambda1548~{\rm\AA}$\ optical depth as a function of wavelength for
a $50~{\rm\AA}$\ sample of spectrum, with one strong \ion{C}{4}
detection.  {\it Middle panel\/}: corresponding \ion{C}{4}
$\lambda1550~{\rm\AA}$\ optical depth shifted to the $\lambda$1548
frame.  {\it Top panel\/}: result of the optical depth selection
criterion of $0.25 < \tau(1548)/\tau(1550) < 1$, confirming the system
as a real \ion{C}{4} doublet.  The optical depth
distributions have been smoothed to $5~{\rm km\ s}^{-1}$, to roughly
Nyquist sample the typical CIV line widths in these spectra.  
Only a small number of points in the
noise satisfy the doublet ratio window and appear as noise spikes in
the cleaned vector of the upper panel.  Below a
wavelength of $5840~{\rm\AA}$ in the lower panel approximately 20\%
of the 409 independent pixels have a raw $\tau (1548)$\ above 0.005. Just
over three quarters of these are eliminated by the doublet ratio
filter, leaving 19 pixels in the cleaned $\tau (1548)$\ vector of the upper
panel which have values above 0.005.  (The dashed line in the top
panel indicates an optical depth of 0.01 to show the scale on the
figure.  It is not a cutoff level.)
}
\label{fig:selec}
\addtolength{\baselineskip}{10pt}
\end{inlinefigure}

%
% Figure 2  Optical depth distribution [TBD]
%
\begin{inlinefigure}
\includegraphics[angle=90,scale=.6]{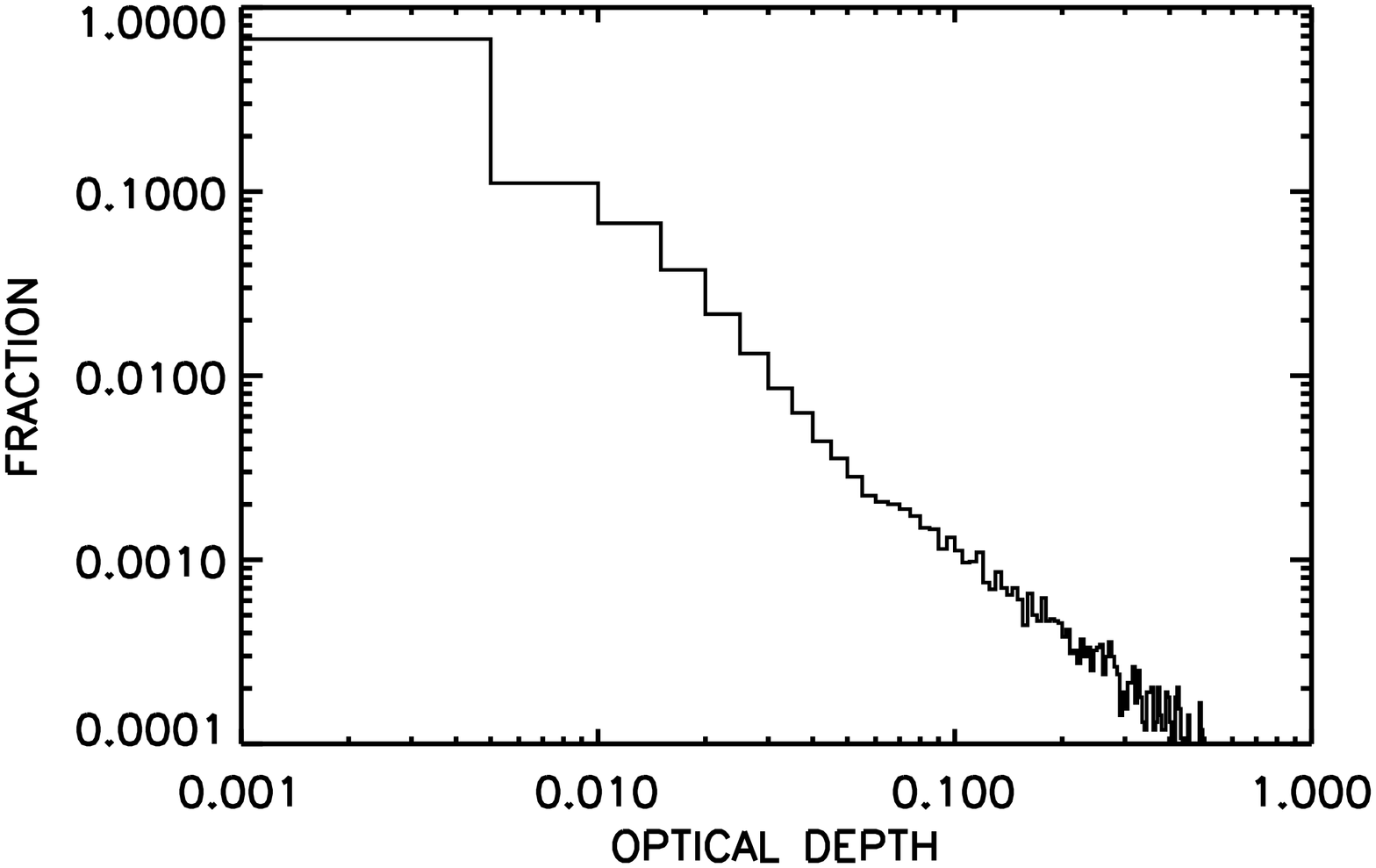}
%\vspace{6pt}
\figurenum{2}
\caption{ Distribution of optical depths over the \ion{C}{4}
wavelength region in the core quasar sample. The result is shown as
the fraction of the total number of pixels lying in 0.005 wide bins in
the optical depth.  Only 3.4\% of the pixels have optical depths above
0.1. Roughly one third of the pixels have optical depths above 0.005.
}
\label{fig:distrib}
\addtolength{\baselineskip}{10pt}
\end{inlinefigure}

%
% Figure 3  Tramlines on
%
\begin{inlinefigure}
\includegraphics[angle=90,scale=.6]{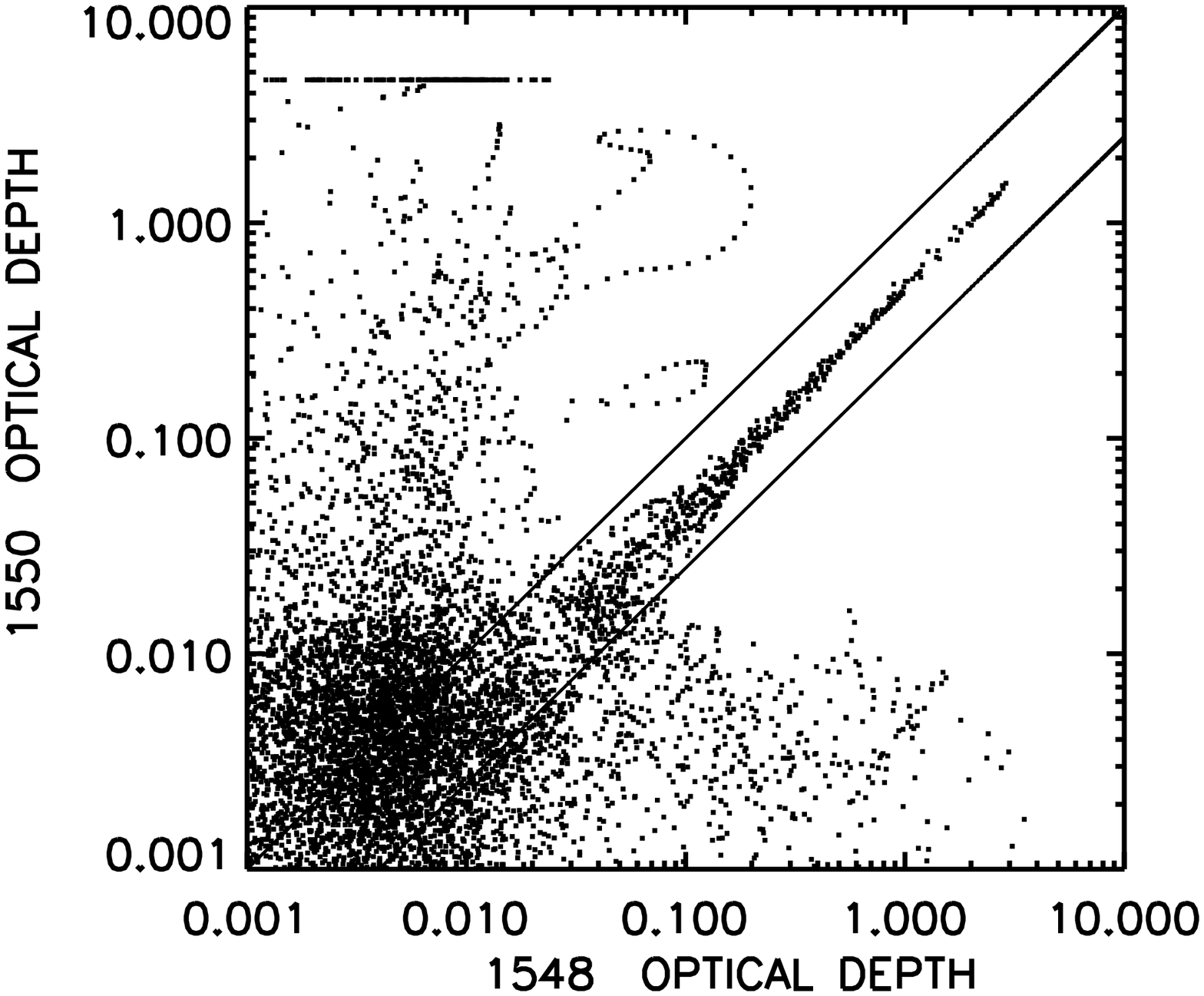}
\vspace{6pt}
\figurenum{3}
\caption{
$\lambda1550~{\rm\AA}$\ optical depth vs
$\lambda1548~{\rm\AA}$\ optical depth  for all the measured optical
depths in HS1946+7658. 
{\it Solid lines\/}: optical depth selection criteria of $0.25 <
\tau(1548)/\tau(1550) < 1$.  Detected \ion{C}{4} doublets are clearly
present visually down to $\tau(1548) \sim 0.01$.  Only positions with
$\tau (1548) < 4$\ are plotted.  Saturated regions appear at $\tau =
4.8$\ in the $\tau (1550)$\ axis.
}
\label{fig:tramon}
\addtolength{\baselineskip}{10pt}
\end{inlinefigure}

%
% Figure 4  Tramlines off
%
\begin{inlinefigure}
\includegraphics[angle=90,scale=.6]{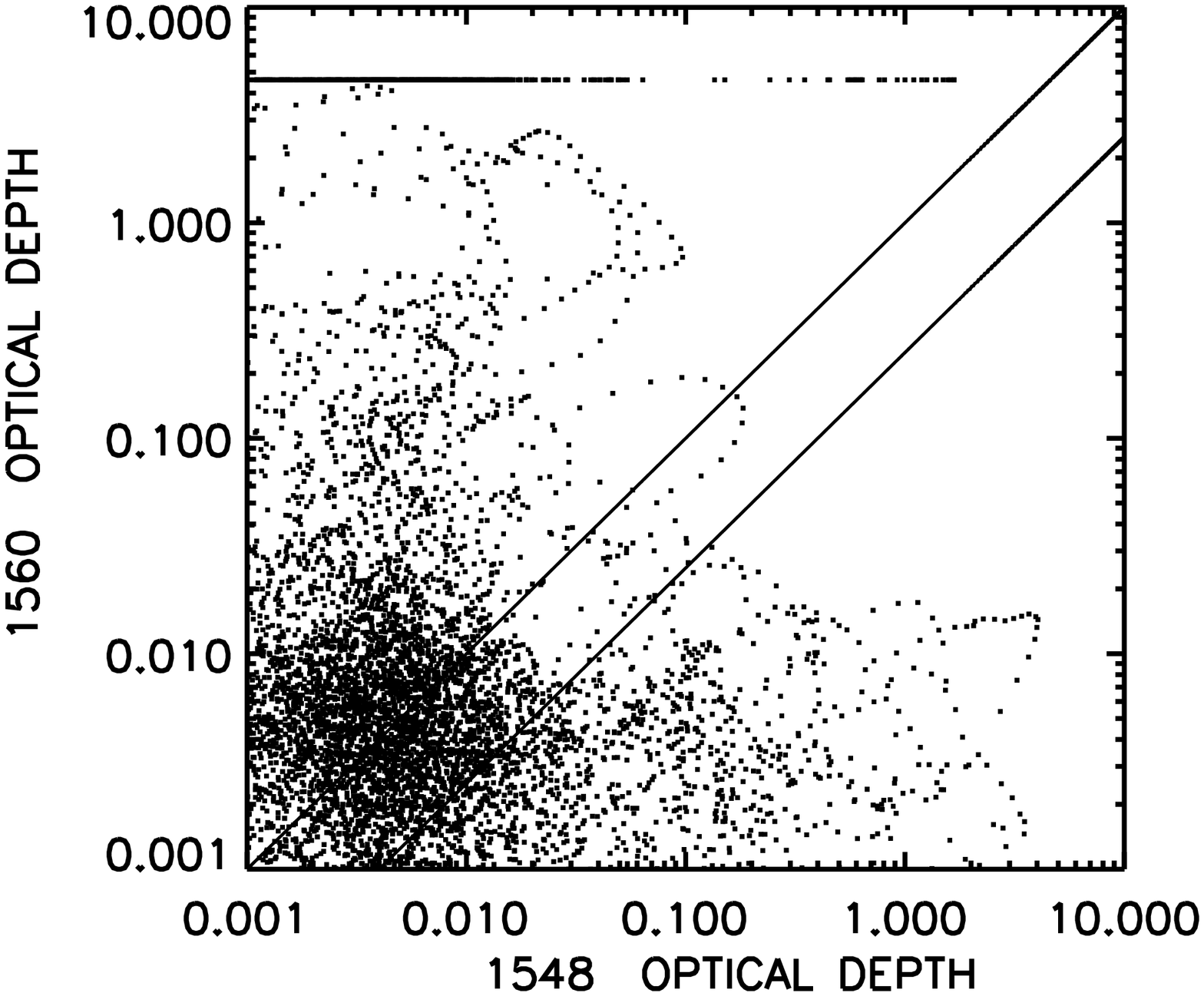}
\vspace{6pt}
\figurenum{4}
\caption{
As in Fig. 2 for artificially generated 'false doublets' with doublet
separation of $12~{\rm\AA}$.  The solid lines have the same
meaning as in Figure~\ref{fig:tramon}.
}
\label{fig:tramoff}
\addtolength{\baselineskip}{10pt}
\end{inlinefigure}

%
% Figure 5  O. d. distribution, on and off
%
\begin{inlinefigure}
\includegraphics[angle=90,scale=.6]{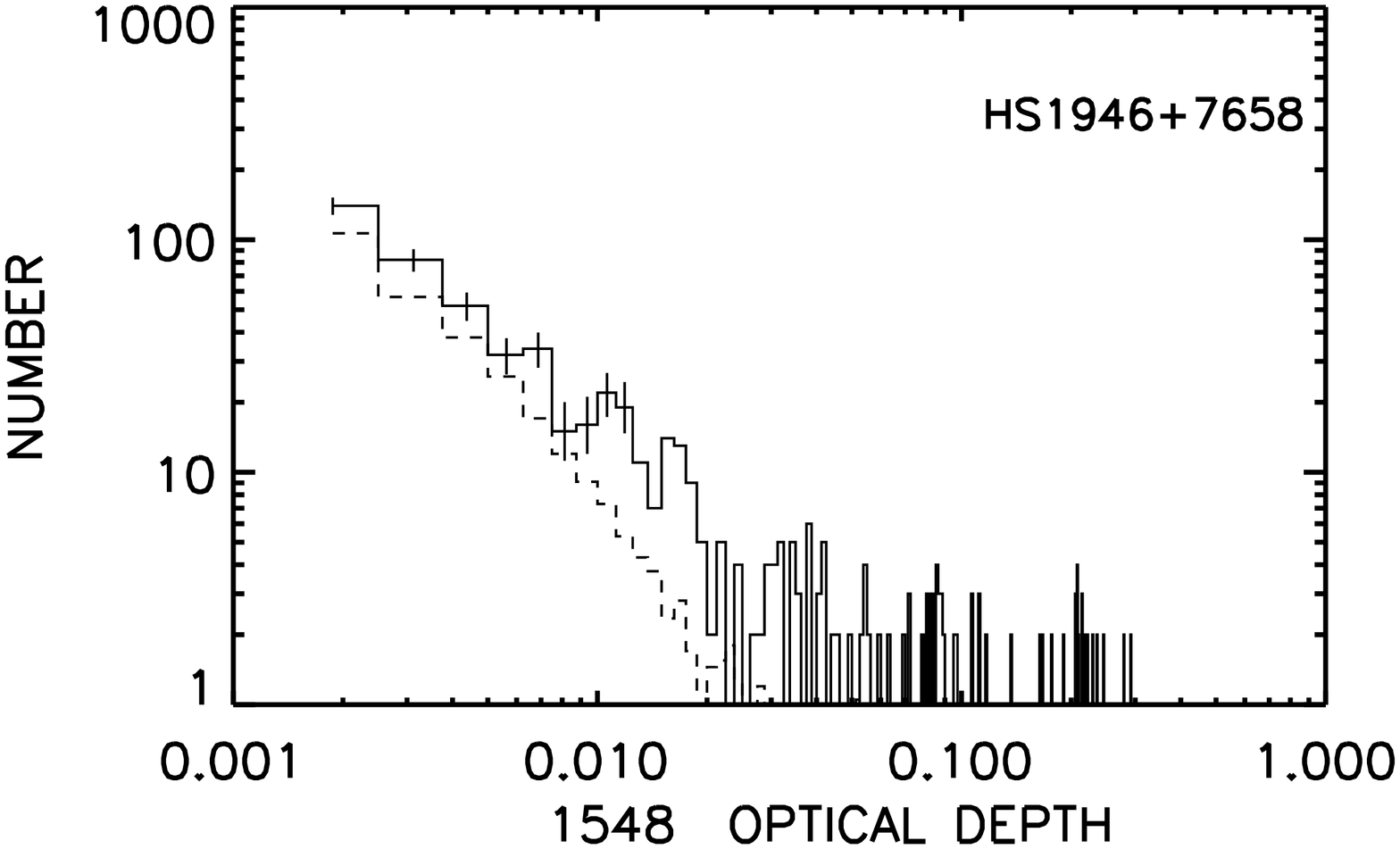}
\vspace{6pt}
\figurenum{5}
\caption{ 
{\it Solid histogram\/}: \ion{C}{4} $\lambda1548~{\rm\AA}$\
optical depth distribution for selected lines in the spectrum of the
$z_{\rm em} = 3.03$\ quasar HS~1946+7658. Error bars are $\pm
1~\sigma$.  Lines were chosen as illustrated in Figs.~\ref{fig:selec}
and \ref{fig:distrib}.  {\it Dashed histogram\/}: average optical
depth distribution retrieved from 20 artificial doublets with
incorrect doublet separations to model contamination of real
\ion{C}{4} doublets by interloper lines, bad pixels, telluric
absorption, etc.  }
\label{fig:odonoff}
\addtolength{\baselineskip}{10pt}
\end{inlinefigure}

%
% Figure 6 O. d. distribution signal
%
\begin{inlinefigure}
\includegraphics[angle=90,scale=.6]{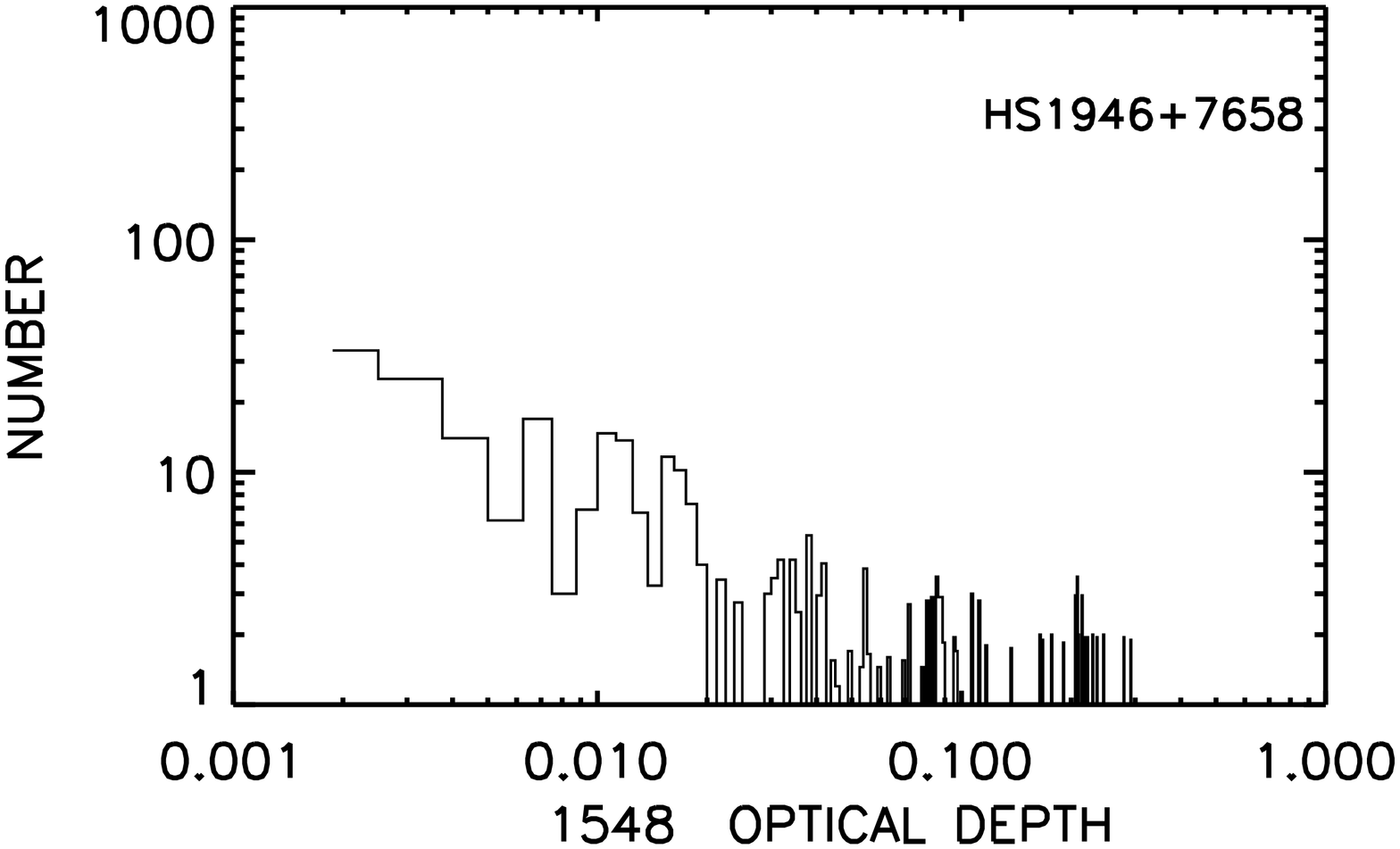}
\vspace{6pt}
\figurenum{6}
\caption{
Difference between the solid and dashed histograms of Figure~\ref{fig:odonoff},
illustrating the \ion{C}{4} $\lambda1548~{\rm\AA}$\ optical depth
distribution statistically corrected for contamination.  There is
significant positive signal down to $\tau(1548) \approx 0.005$.
}
\label{fig:odsignal}
\addtolength{\baselineskip}{10pt}
\end{inlinefigure}

%
% Figure 7 Line recovery 
%
\begin{inlinefigure}
\includegraphics[angle=90,scale=.6]{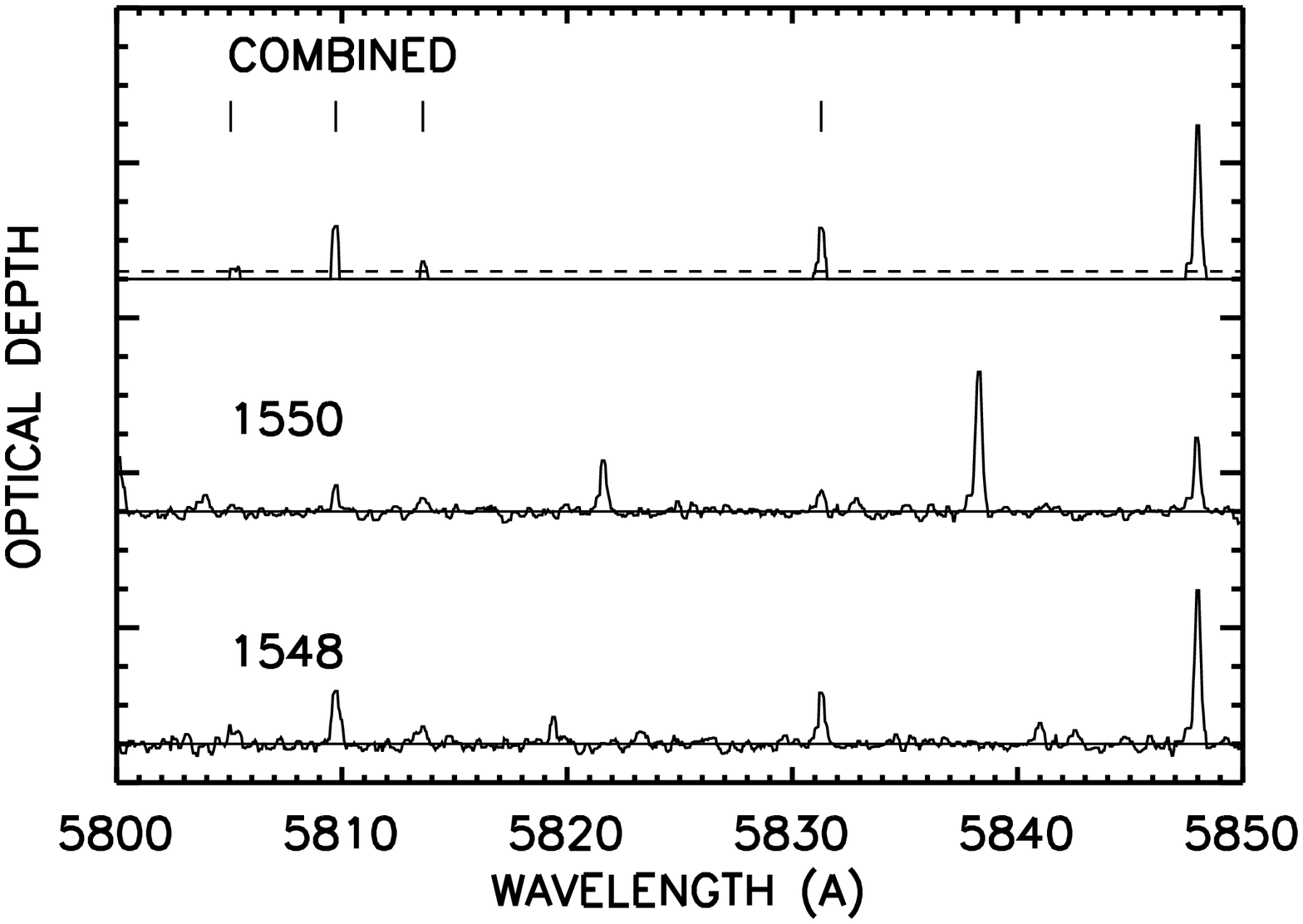}
\vspace{6pt}
\figurenum{7}
\caption{ 
Illustration of the method of measuring the recovery rate of
artificial doublets.  Two artifical \ion{C}{4} doublets with column
density $\log N({\rm C~IV}) = 11.75$\ and two with $\log N({\rm C~IV})
= 12.75$\ have been inserted into the region of the spectrum of
HS1946+7658 shown in Figure~\ref{fig:selec}.  The panels have the same
meaning as in Fig.~\ref{fig:selec}.  Short vertical lines in the
top panel show the wavelengths of the introduced artificial lines.  }
\label{fig:recovery}
\addtolength{\baselineskip}{10pt}
\end{inlinefigure}

%
% Figure 8 Line recovery rate
%
\begin{inlinefigure}
\includegraphics[angle=90,scale=.6]{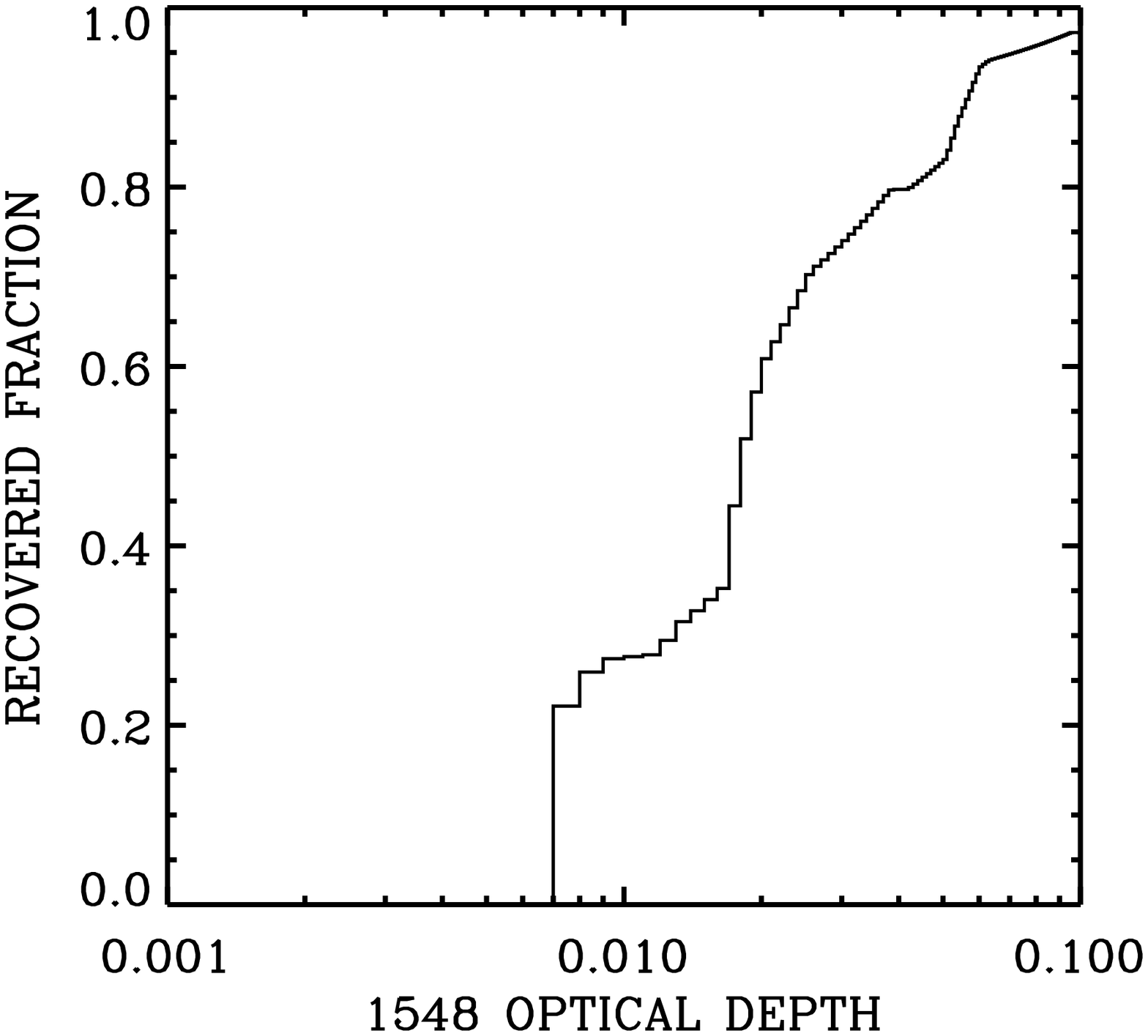}
\vspace{6pt}
\figurenum{8}
\caption{
Fraction of recovered pixels as a function of the input
optical depth in the $1548~{\rm\AA}$ line for the 5 quasars in the
core quasar sample. At higher optical depths the small incompleteness
is primarily caused by line blending. The recovery begins to drop
rapidly at lower optical depths because of noise scattering which
moves ratios outside the doublet window. The recovery in these quasars
drops to 0.5 at $\tau=0.017$ and to 0.25 at $\tau=0.007$, at which
point we consider we have reached the useful limit of the data.
}
\label{fig:rate}
\addtolength{\baselineskip}{10pt}
\end{inlinefigure}

%
% Figure 9 Pseudo cloud example
%
\begin{inlinefigure}
\includegraphics[angle=90,scale=.6]{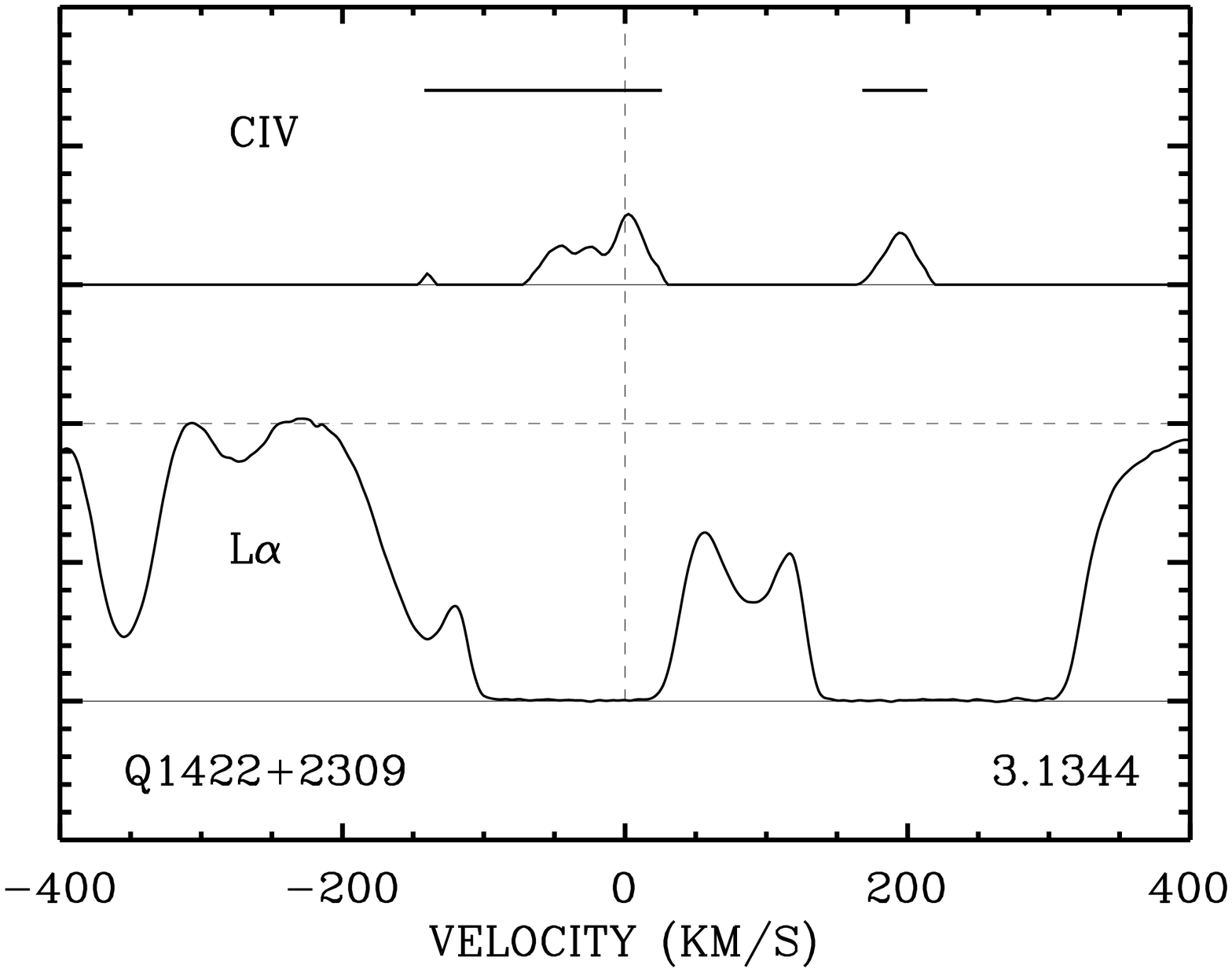}
\vspace{6pt}
\figurenum{9}
\caption{ 
Illustration of the generation of cloud complexes in a
region of the spectrum of the quasar Q1422+2309. {\it Top panel\/}:
cleaned \ion{C}{4} optical depth vector. The analysis, with a
$100~{\rm km\ s}^{-1}$\ cloud window, splits this region into two
complexes that are marked by the horizontal solid lines. {\it Lower
panel\/}: corresponding region of the Ly$\alpha$ line, which shows
that the complex breakdown matches well to the Ly$\alpha$
structure. However, the exact breakdown is subject to the exact choice
of window. A $150~{\rm km\ s}^{-1}$ window would reduce this
particular system to one complex whereas a $50~{\rm km\ s}^{-1}$window
would split it into three parts.  
}
\label{fig:pseudoex}
\addtolength{\baselineskip}{10pt}
\end{inlinefigure}

%
% Figure 10 Effect of window size on complexes
%
\begin{inlinefigure}
\includegraphics[angle=90,scale=.6]{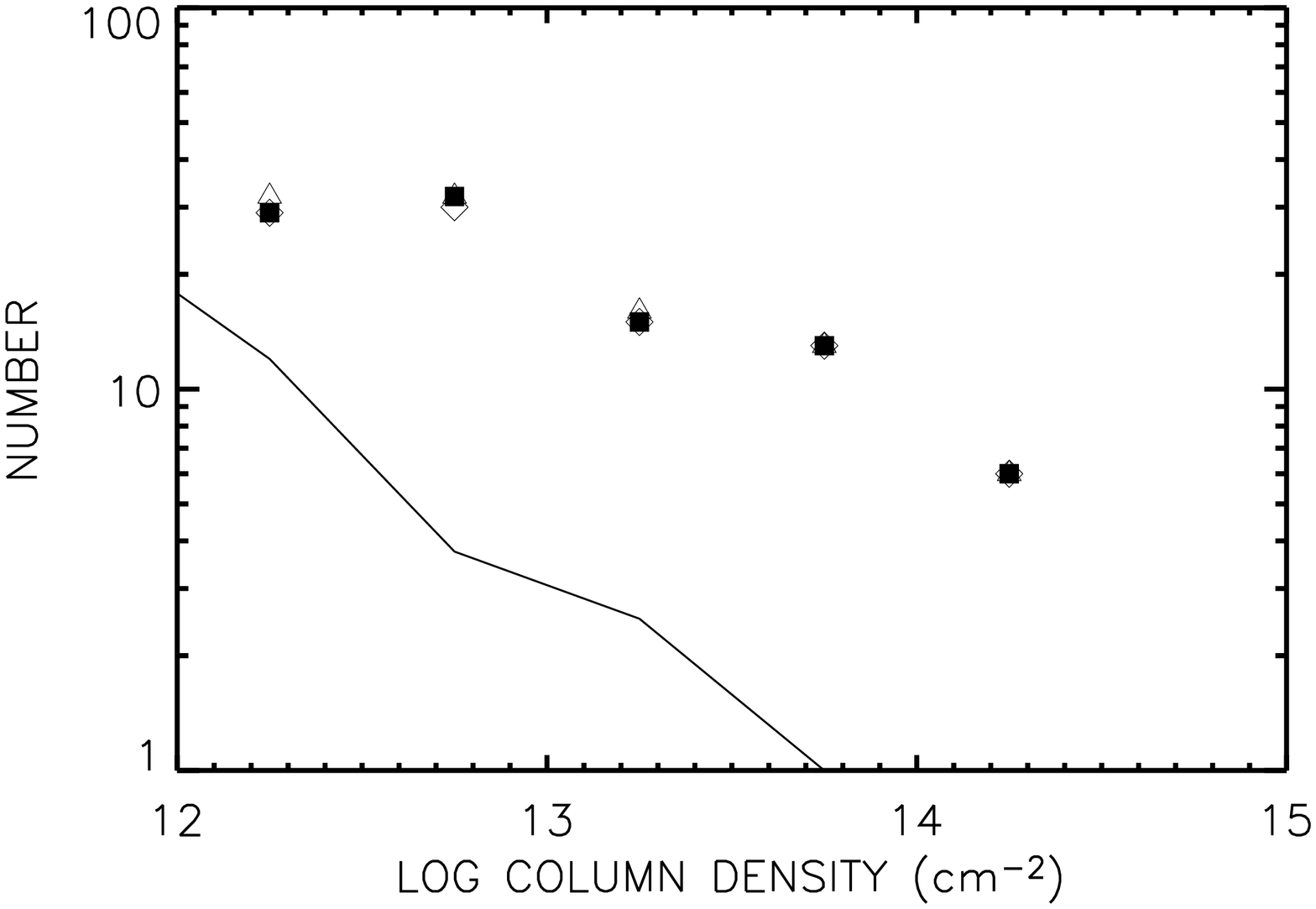}
\vspace{6pt}
\figurenum{10}
\caption{ 
Number of complexes retrieved in the core quasar sample as a
function of column density for three velociy windows: $75~{\rm km\
s}^{-1}$ ({\it triangles}), $150~{\rm km\ s}^{-1}$ ({\it squares}),
and $300~{\rm km\ s}^{-1}$ ({\it diamonds}). The effect of widening
the window is to blend weaker systems into larger complexes and so
reduce the number of low column density sytems but, as can be seen
from the figure, the effect is not large for reasonable values of the
velocity window. {\it Solid line\/}: number of systems found as a
function of column density for a set of artificial doublets
approximating the C~IV separation; this shows the degree of
contamination at a given column density.  
}
\label{fig:window}
\addtolength{\baselineskip}{10pt}
\end{inlinefigure}

%
% Figure 11 CIV distn from psuedo-clouds
%
\begin{inlinefigure}
\includegraphics[angle=90,scale=.6]{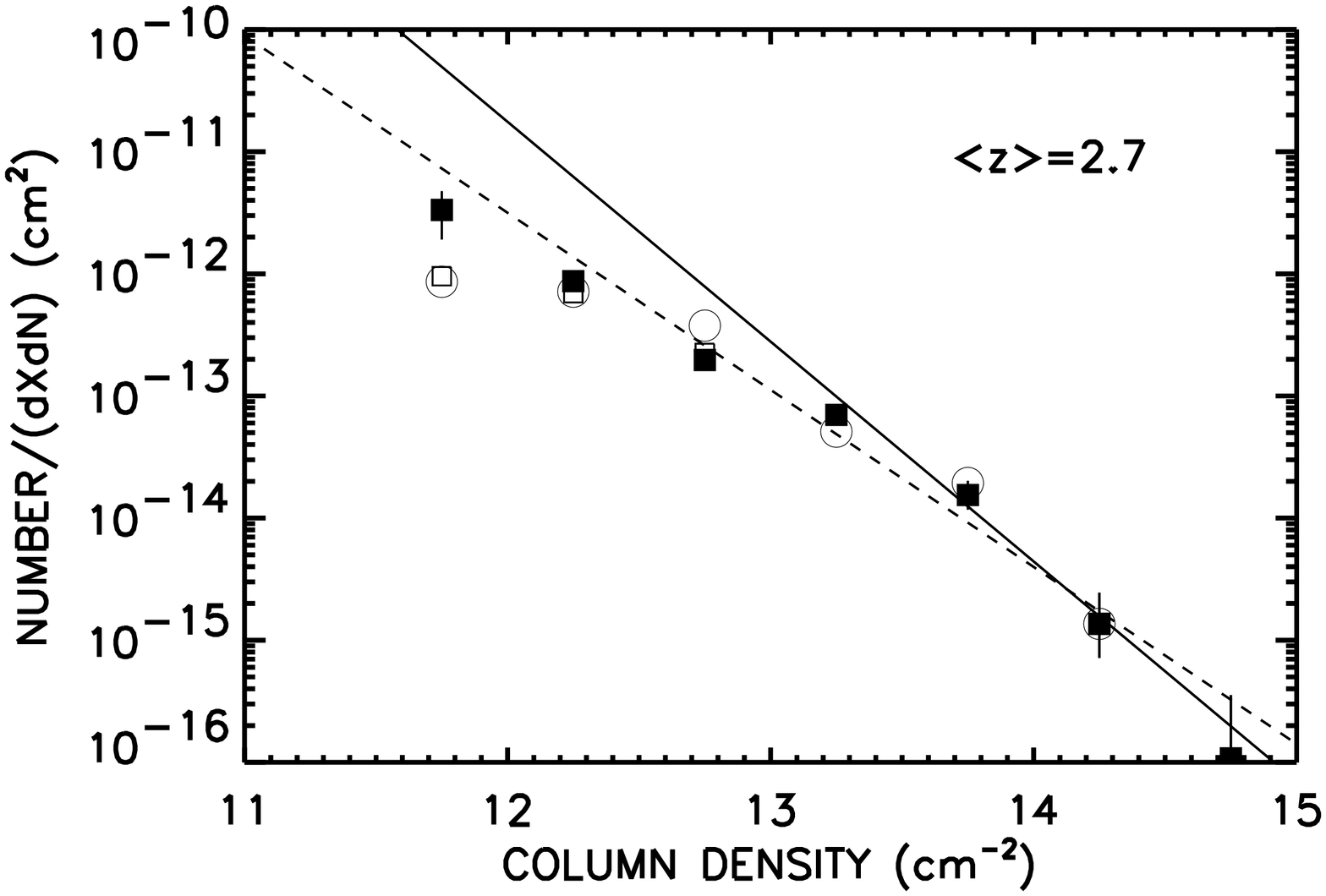}
\vspace{6pt}
\figurenum{11}
\caption{ 
{\it Open squares\/}: \ion{C}{4} column density distribution
of pseudo-clouds determined using a velocity window of $100~{\rm km\
s}^{-1}$ (see text) from the \ion{C}{4} optical depths retrieved by the
superPOD method from the core quasar sample.  The average \ion{C}{4}
redshift is 2.7.  {\it Filled squares\/}: distribution corrected for
incompleteness, with $1~\sigma$ error bars.  {\it Open circles\/}:
distribution obtained by Voigt profile fitting of \ion{C}{4} doublets
selected by hand from the spectra, with no incompleteness correction.
{\it Dotted line\/}: fit to the data with power law index $-1.44$.
{\it Solid line\/}: power law slope $= -1.7$, which fits the $N({\rm
C~IV}) > 10^{13}~{\rm cm}^{-2}$\ sample.  
}
\label{fig:c4dist}
\addtolength{\baselineskip}{10pt}
\end{inlinefigure}

%
% Figure 12 Omega(CIV) from pseudo-clouds
%
\begin{inlinefigure}
\includegraphics[angle=90,scale=.6]{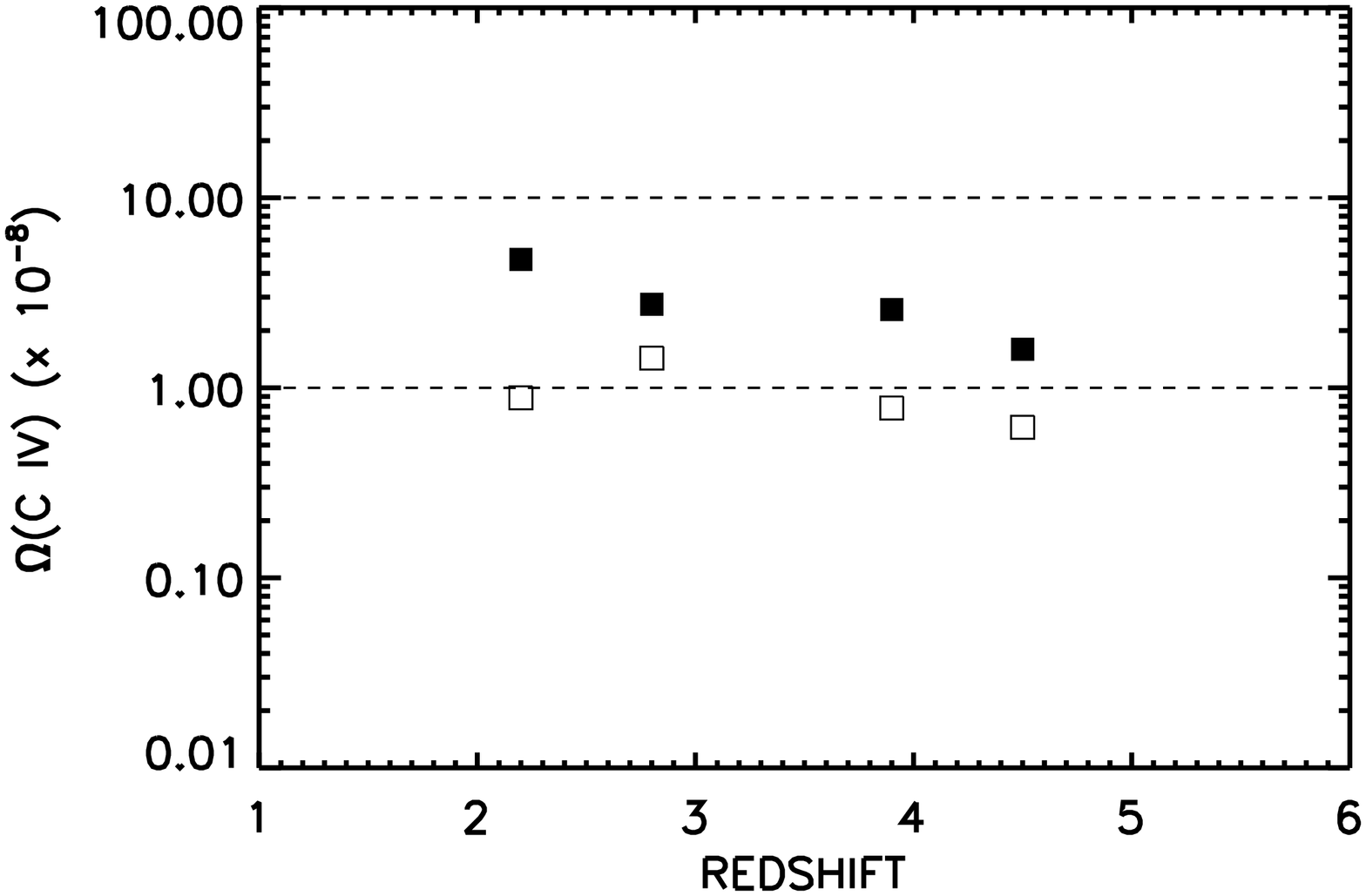}
\vspace{6pt}
\figurenum{12}
\caption{
{\it Filled squares\/}: $\Omega({\rm C IV})$\ as a function of
redshift computed from \ion{C}{4} column density distributions of
pseudo-clouds with $12 < \log N({\rm C IV}) < 15$\ obtained from
\ion{C}{4} optical depths retrieved by the superPOD method from
various samples of quasars with $ 2 < z < 5.5$.
{\it Open squares\/}: as above, with $13 < \log N({\rm C IV})
< 14$.
}
\label{fig:c4omega}
\addtolength{\baselineskip}{10pt}
\end{inlinefigure}

%
% Figure 13 CII/CIV vs z
%
\begin{inlinefigure}
\includegraphics[angle=90,scale=.6]{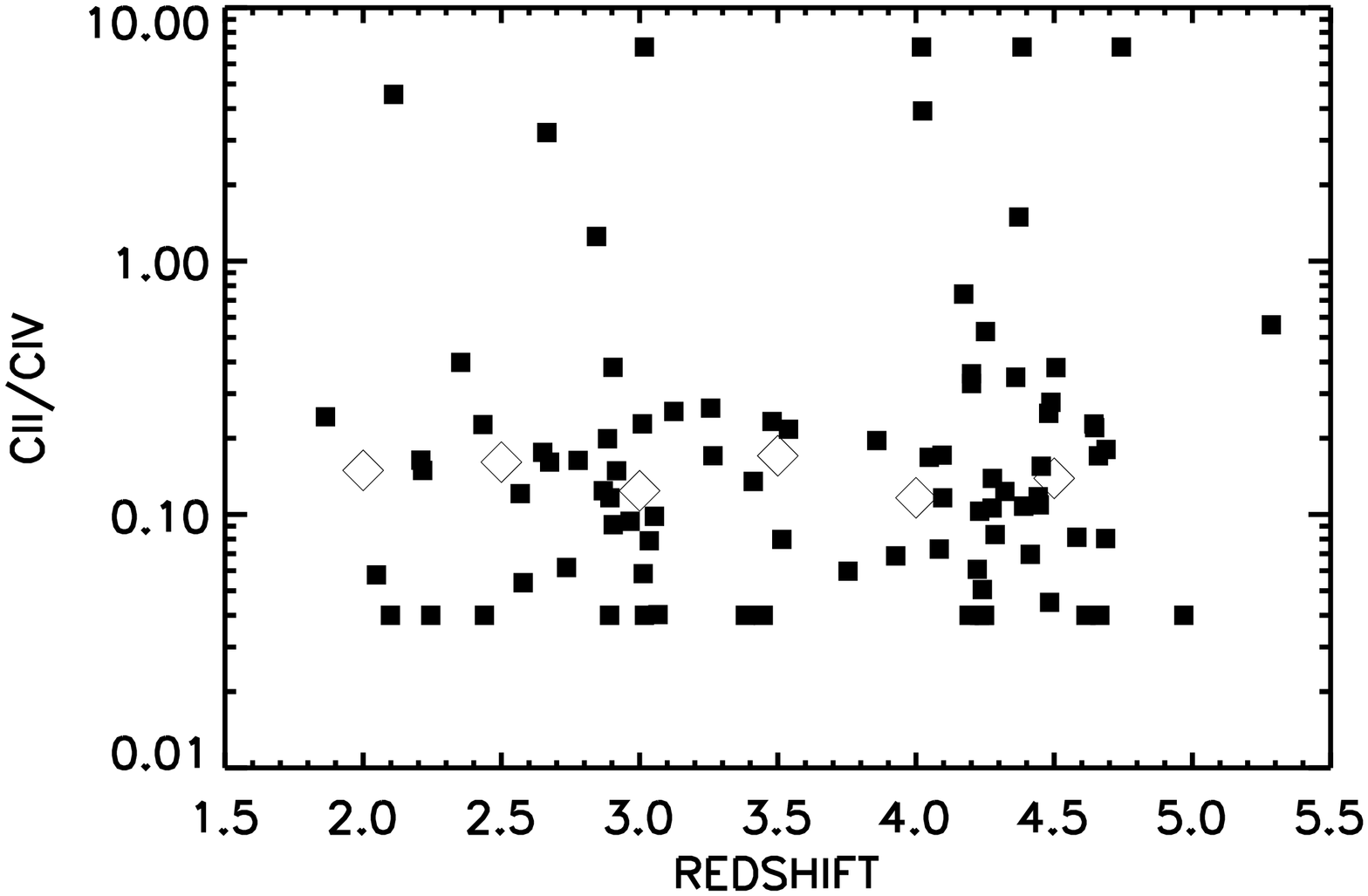}
\vspace{6pt}
\figurenum{13}
\caption{
{\it Small filled squares\/}: \ion{C}{2}/\ion{C}{4} as a function of
redshift determined from the integrated strengths of \ion{C}{2} and
\ion{C}{4} in \ion{C}{4}-selected pseudo-clouds (see text) with a
width of $100~{\rm km\ s}^{-1}$.  \ion{C}{2} optical depth vectors
have not been cleaned to remove contamination.  {\it Large open
diamonds\/}:  median value of \ion{C}{2}/\ion{C}{4} in each redshift
bin, calculated directly from \ion{C}{2} and \ion{C}{4} optical depth
vectors. 
}
\label{fig:c2c4}
\addtolength{\baselineskip}{10pt}
\end{inlinefigure}

%
% Figure 14 SiIV/CIV vs z
%
\begin{inlinefigure}
\includegraphics[angle=90,scale=.6]{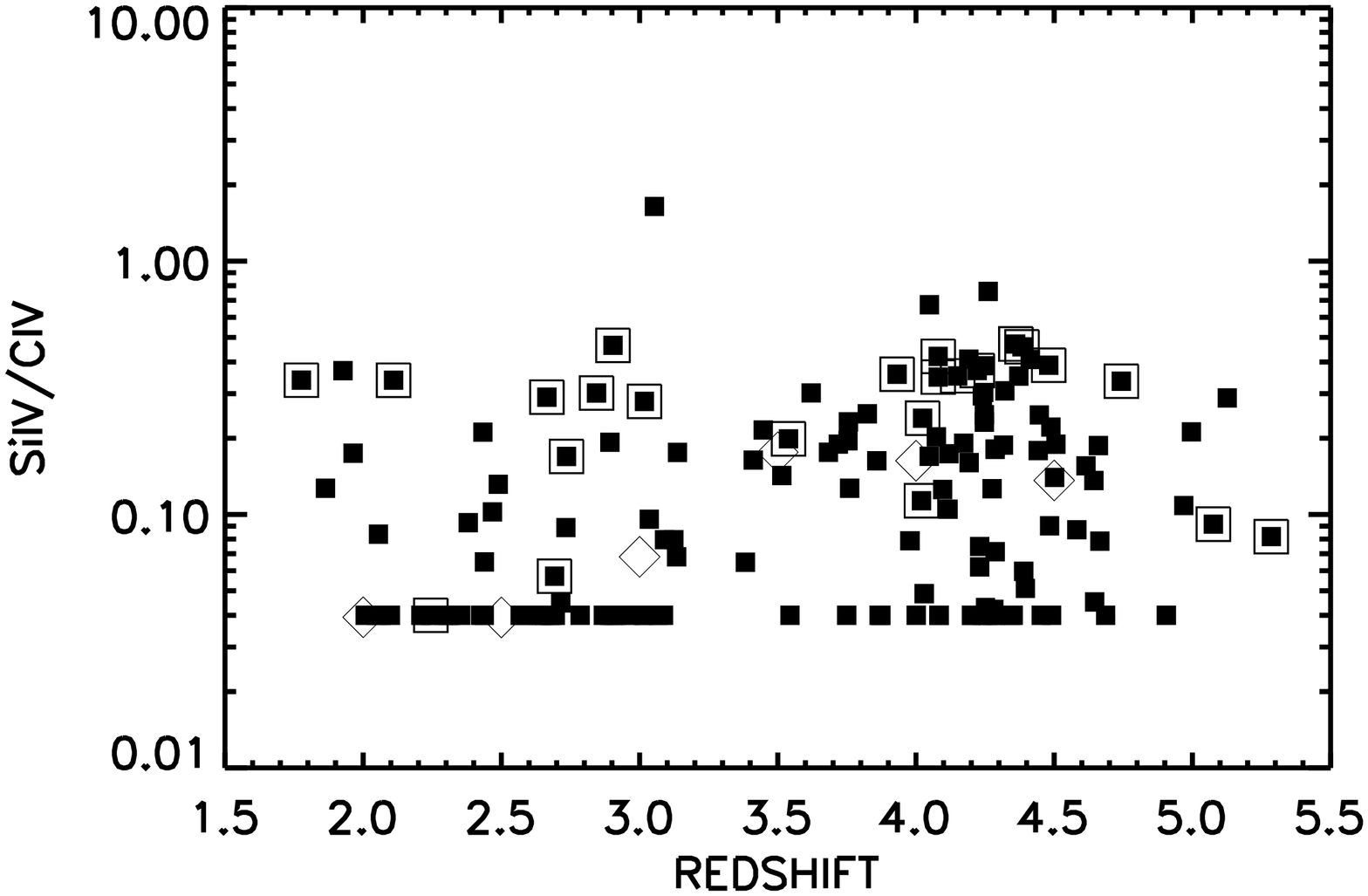}
\vspace{6pt}
\figurenum{14}
\caption{ 
{\it Small filled squares\/}: \ion{Si}{4}/\ion{C}{4} as a
function of redshift determined from the integrated strengths of
\ion{Si}{4} and \ion{C}{4} in \ion{C}{4}-selected pseudo-clouds (see
text) with a width of $100~{\rm km\ s}^{-1}$.  Both \ion{Si}{4} and
\ion{C}{4} optical depths have been cleaned to remove contamination
(see text).  {\it Large open diamonds\/}: median value of
\ion{Si}{4}/\ion{C}{4} in each redshift bin, calculated directly from
\ion{Si}{4} and \ion{C}{4} optical depth vectors.  {\it Large open
squares\/}: positions of strong absorbers with $N({\rm C~IV})$\ and
$N({\rm C~II}) > 10^{14}~{\rm cm}^{-2}$. 
}
\label{fig:si4c4}
\addtolength{\baselineskip}{10pt}
\end{inlinefigure}

%
% Figure 15 SiIV/CIV KS test
%
\begin{inlinefigure}
\includegraphics[angle=90,scale=.6]{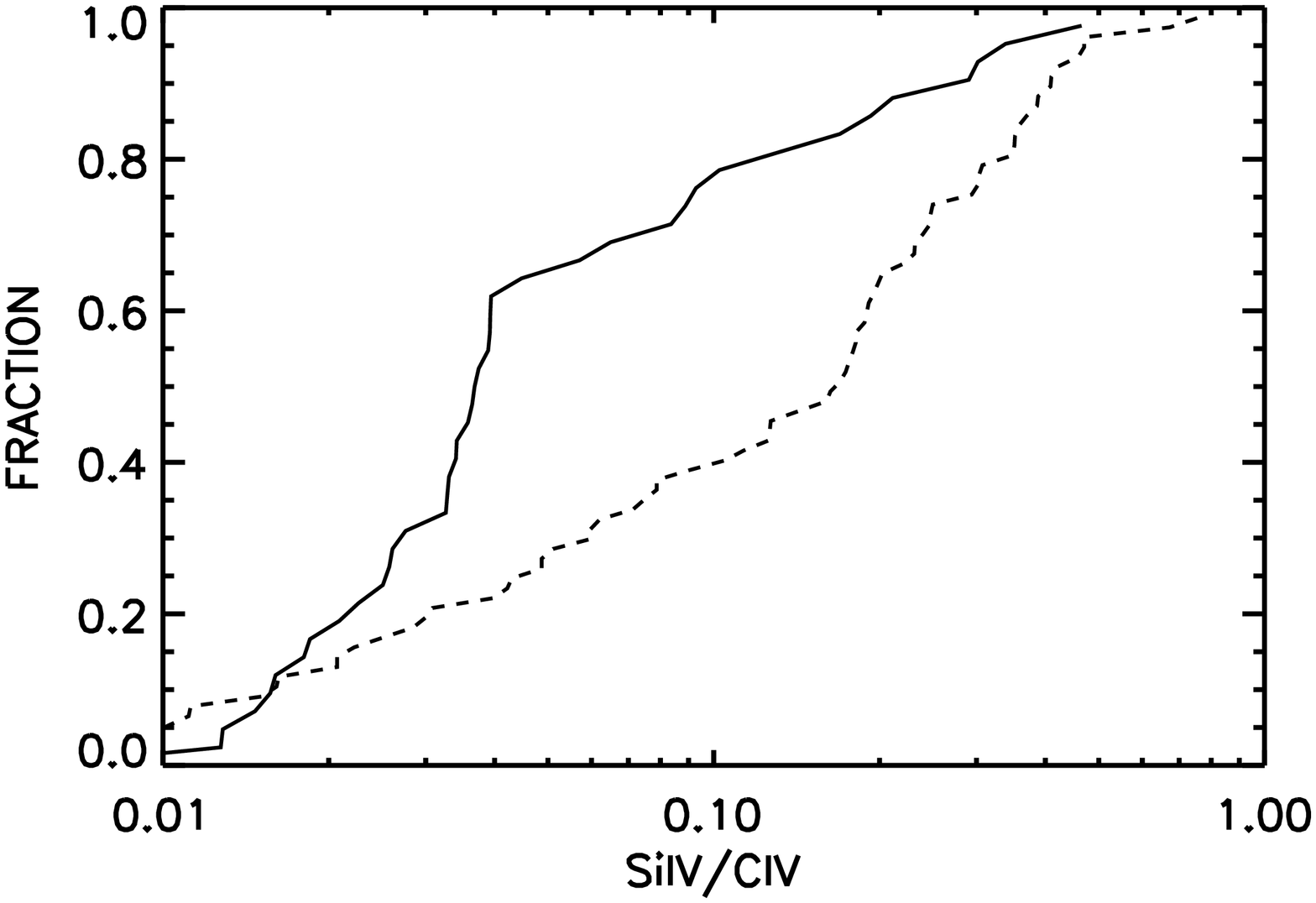}
\vspace{6pt}
\figurenum{15}
\caption{
Kolmogorov-Smirnov test of the hypothesis that the
\ion{Si}{4}/\ion{C}{4} ratio in the cloud complexes at $3.5 < z <
4.5$\ ({\it dashed line}) is the same as at $2 < z < 3$\ ({\it solid
line}).  The  \ion{Si}{4}/\ion{C}{4} ratios are significantly weighted
to higher values at the the higher redshifts despite the fact that the
distribution of \ion{C}{4} column densities has not changed.
}
\label{fig:ks}
\addtolength{\baselineskip}{10pt}
\end{inlinefigure}

%
% Figure 16a o.d. distribution vs column density (a) HIRES
%
\begin{inlinefigure}
\includegraphics[angle=90,scale=.6]{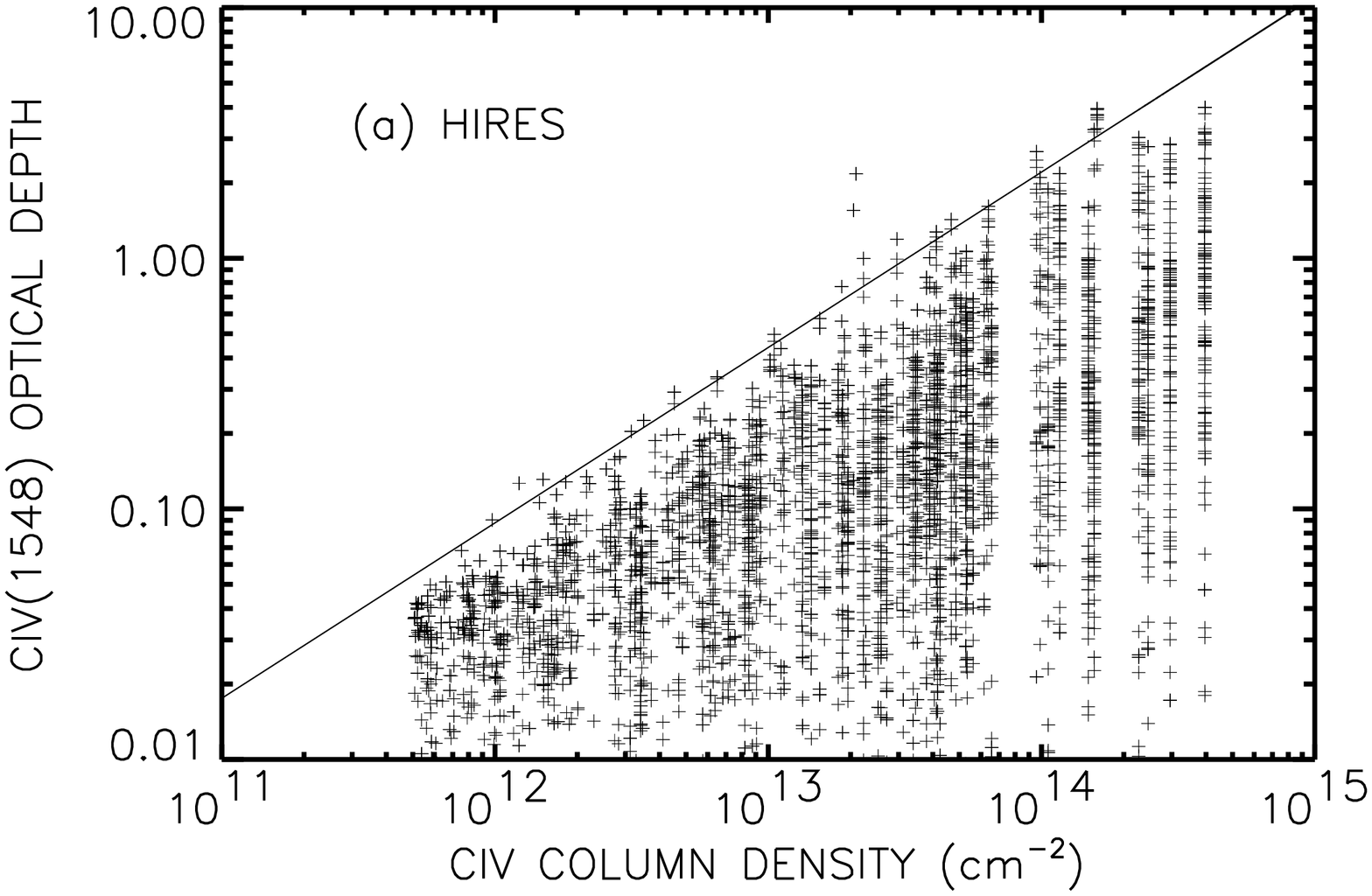}
\vspace{6pt}
\figurenum{16a}
\caption{
Cleaned \ion{C}{4}(1548) optical depths versus the column density of the
complex in which they lie.  HIRES data are shown in panel (a) and ESI
data in panel (b).  For the HIRES data, there is a tight relation,
$\tau \sim N^{0.7}$\ ({\it solid line}) between the peak optical depth
and the column density of the complex.  This is washed out in the
lower resolution ESI data.
}
\label{fig16a}
\addtolength{\baselineskip}{10pt}
\end{inlinefigure}

%
% Figure 16b o.d. distribution vs column density (b) ESI
%
\begin{inlinefigure}
\includegraphics[angle=90,scale=.6]{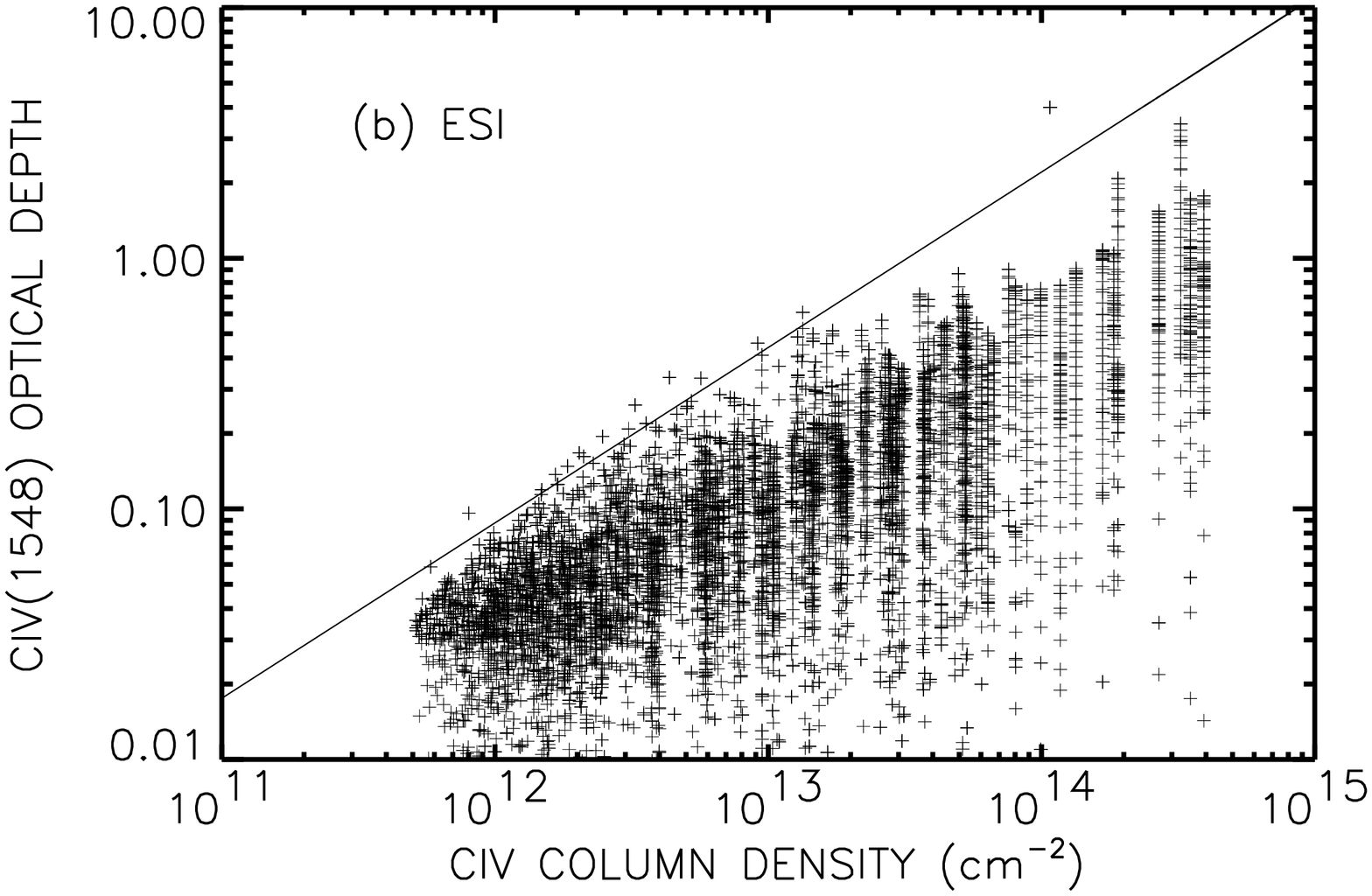}
\vspace{6pt}
\figurenum{16b}
\caption{ 
}
\label{fig16b}
\addtolength{\baselineskip}{10pt}
\end{inlinefigure}

%
% Figure 17 Omega(ion) from optical depths
%
\begin{inlinefigure}
\includegraphics[angle=90,scale=.6]{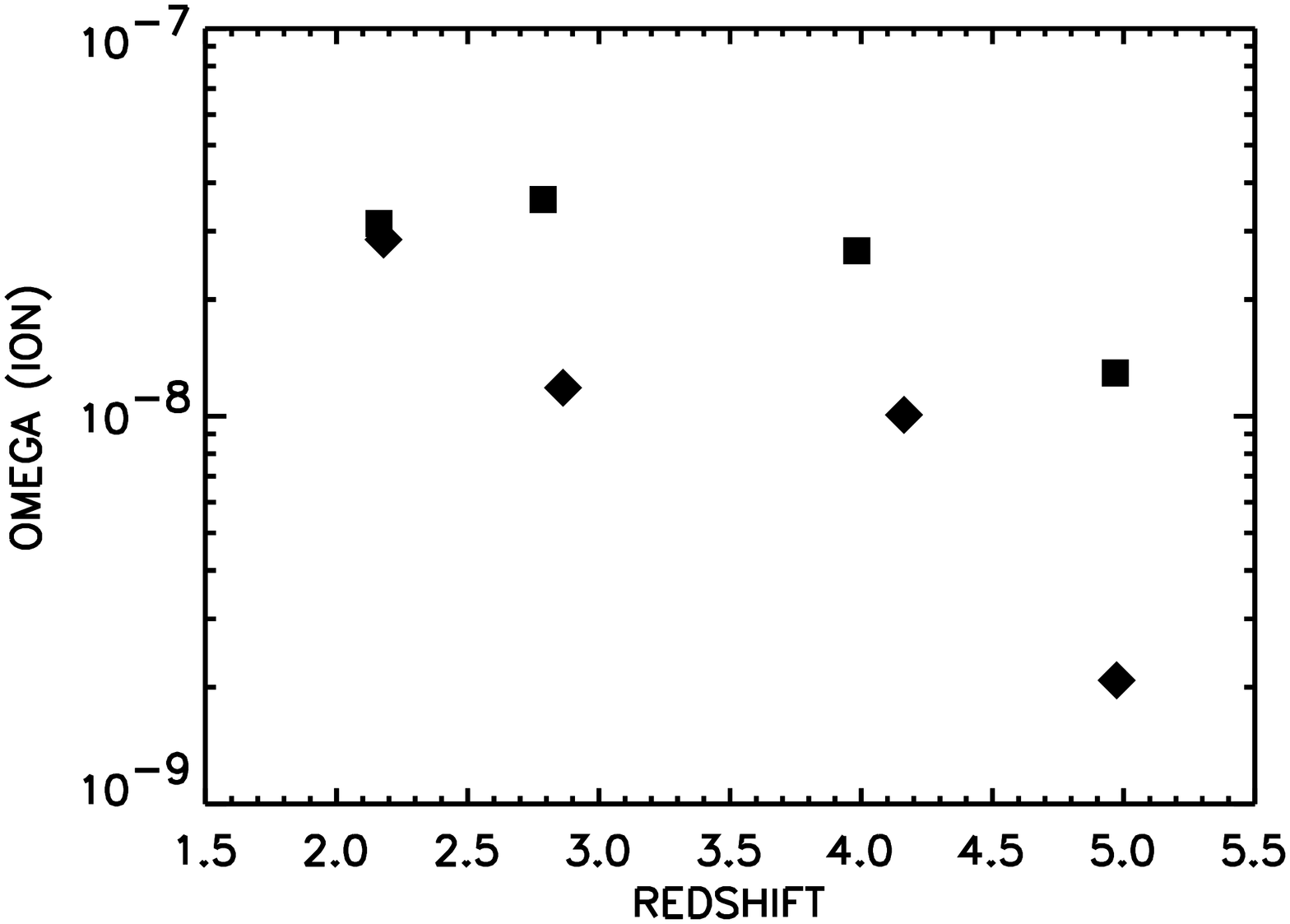}
\vspace{6pt}
\figurenum{17}
\caption{
{\it Filled squares\/}: average $\Omega({\rm C IV})$\ as a function of
redshift computed directly from \ion{C}{4} optical depths retrieved by
the superPOD method from various samples of quasars with $ 2 < z <
5.5$.  Redshift bins are [1.5,2.5], [2.5,3.5], [3.5,4.5], [4.5,5.5],
with the observation shown at the mean redshift in the interval.
{\it Filled diamonds\/}:  as above, for \ion{Si}{4}.
}
\label{fig:odomega}
\addtolength{\baselineskip}{10pt}
\end{inlinefigure}

%
% Figure 18a CIV o.d. distributions
%
\begin{inlinefigure}
\includegraphics[angle=90,scale=.6]{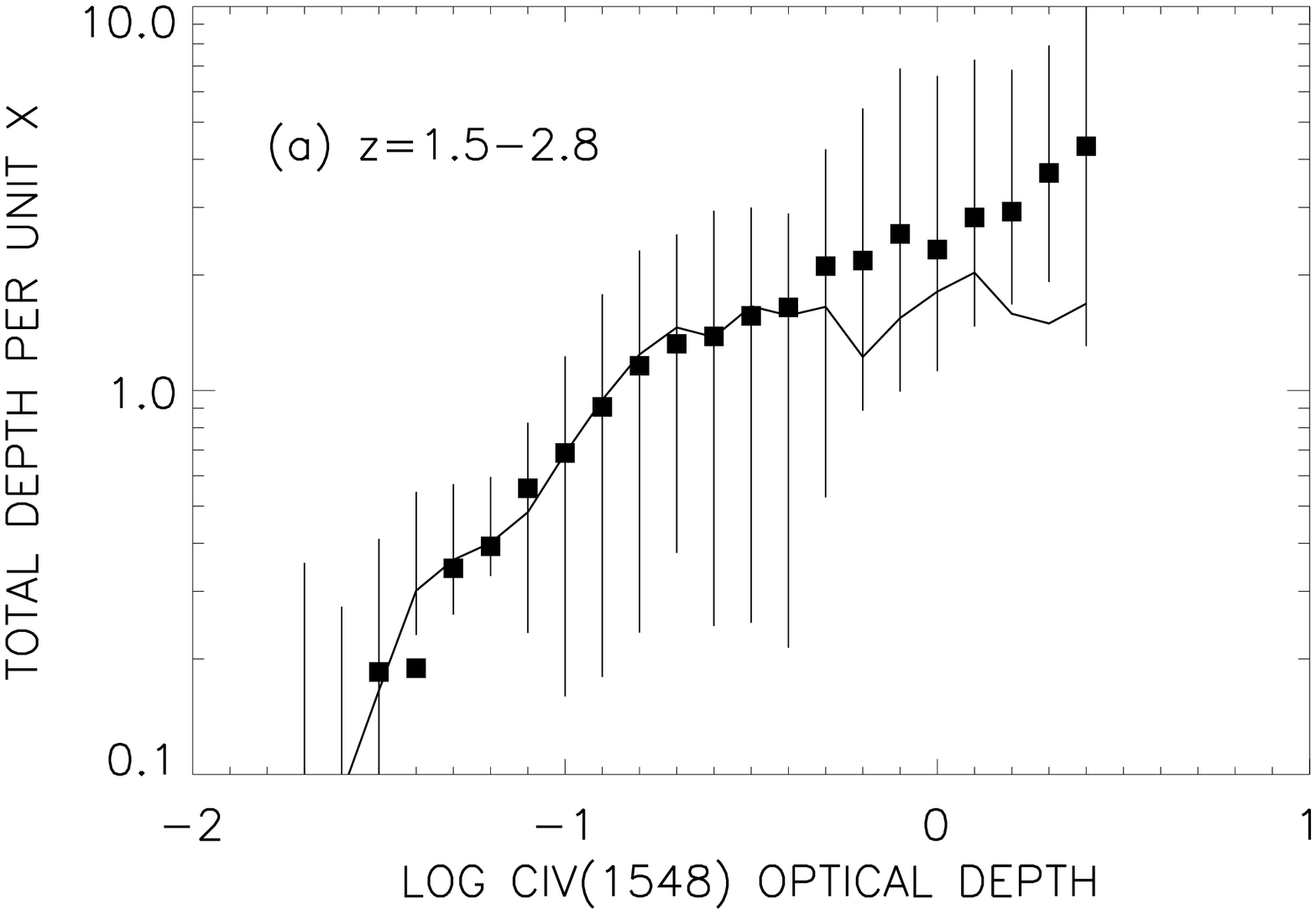}
\vspace{6pt}
\figurenum{18a}
\caption{
Total absorption (the sum of the optical depths) per unit x in optical
depth bins of 0.1.  Each panel contains five quasars ordered by
redshift.  The results in (a) and (b) are based on HIRES data, and
those in (c), (d) and (e) are based on ESI data.  {\it Solid line\/}:
distribution of the core sample.  In (c), (d) and (e) this is
smoothed to the ESI spectral resolution.
}
\label{fig18a}
\addtolength{\baselineskip}{10pt}
\end{inlinefigure}

%
% Figure 18b CIV o.d. distributions
%
\begin{inlinefigure}
\includegraphics[angle=90,scale=.6]{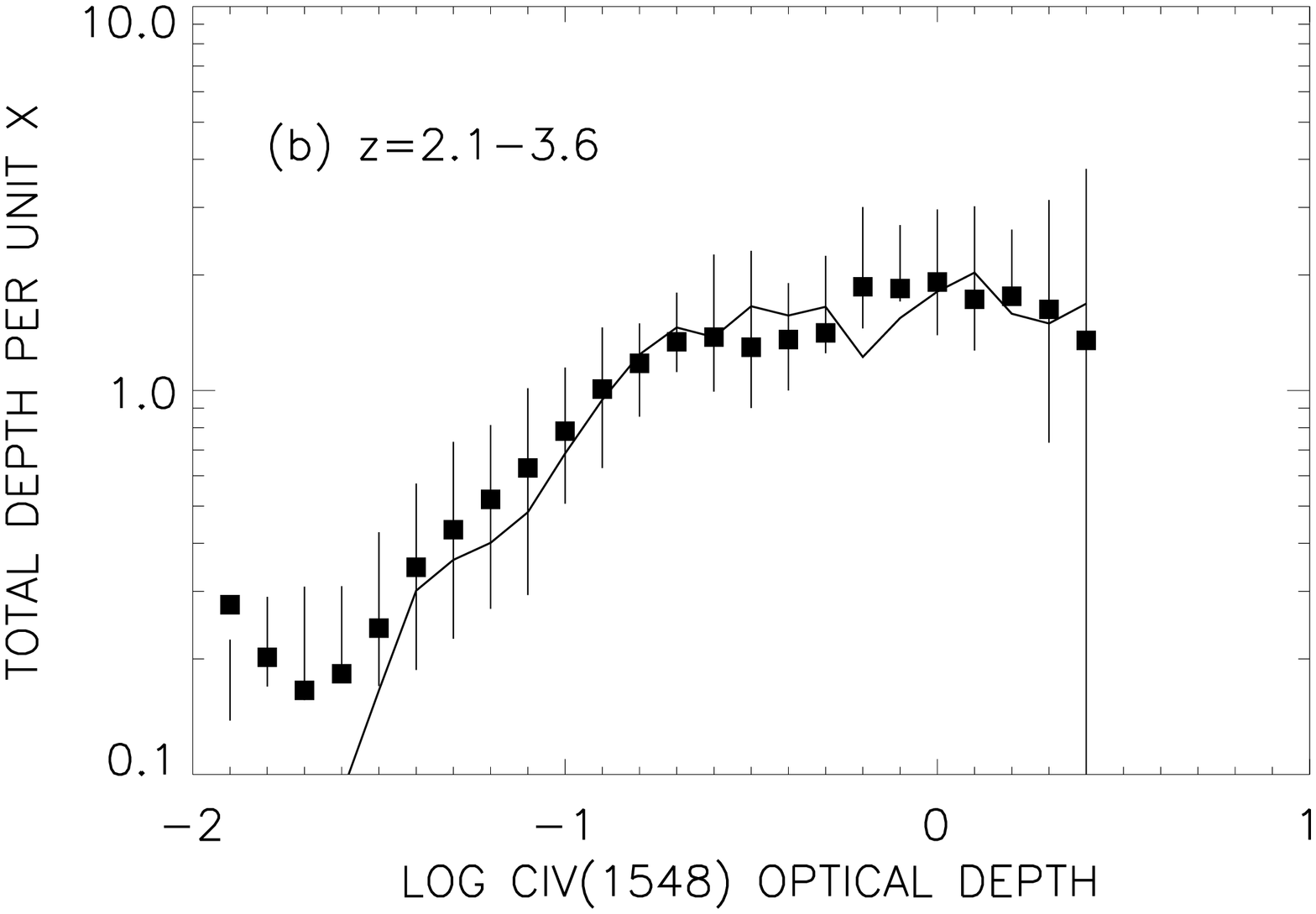}
\vspace{6pt}
\figurenum{18b}
\caption{
}
\label{fig18b}
\addtolength{\baselineskip}{10pt}
\end{inlinefigure}

%
% Figure 18c CIV o.d. distributions
%
\begin{inlinefigure}
\includegraphics[angle=90,scale=.6]{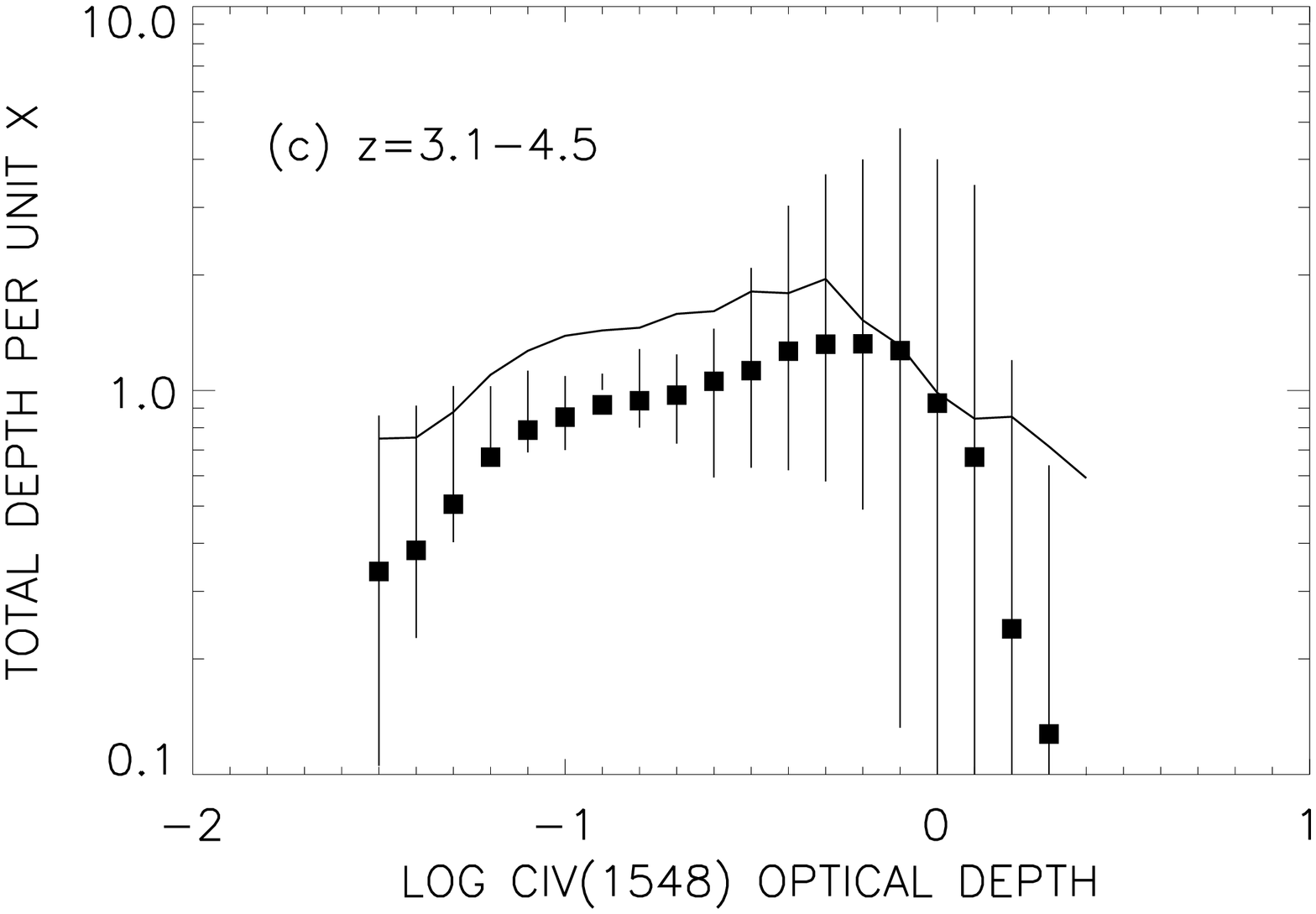}
\vspace{6pt}
\figurenum{18c}
\caption{
}
\label{fig18c}
\addtolength{\baselineskip}{10pt}
\end{inlinefigure}

%
% Figure 18d CIV o.d. distributions
%
\begin{inlinefigure}
\includegraphics[angle=90,scale=.6]{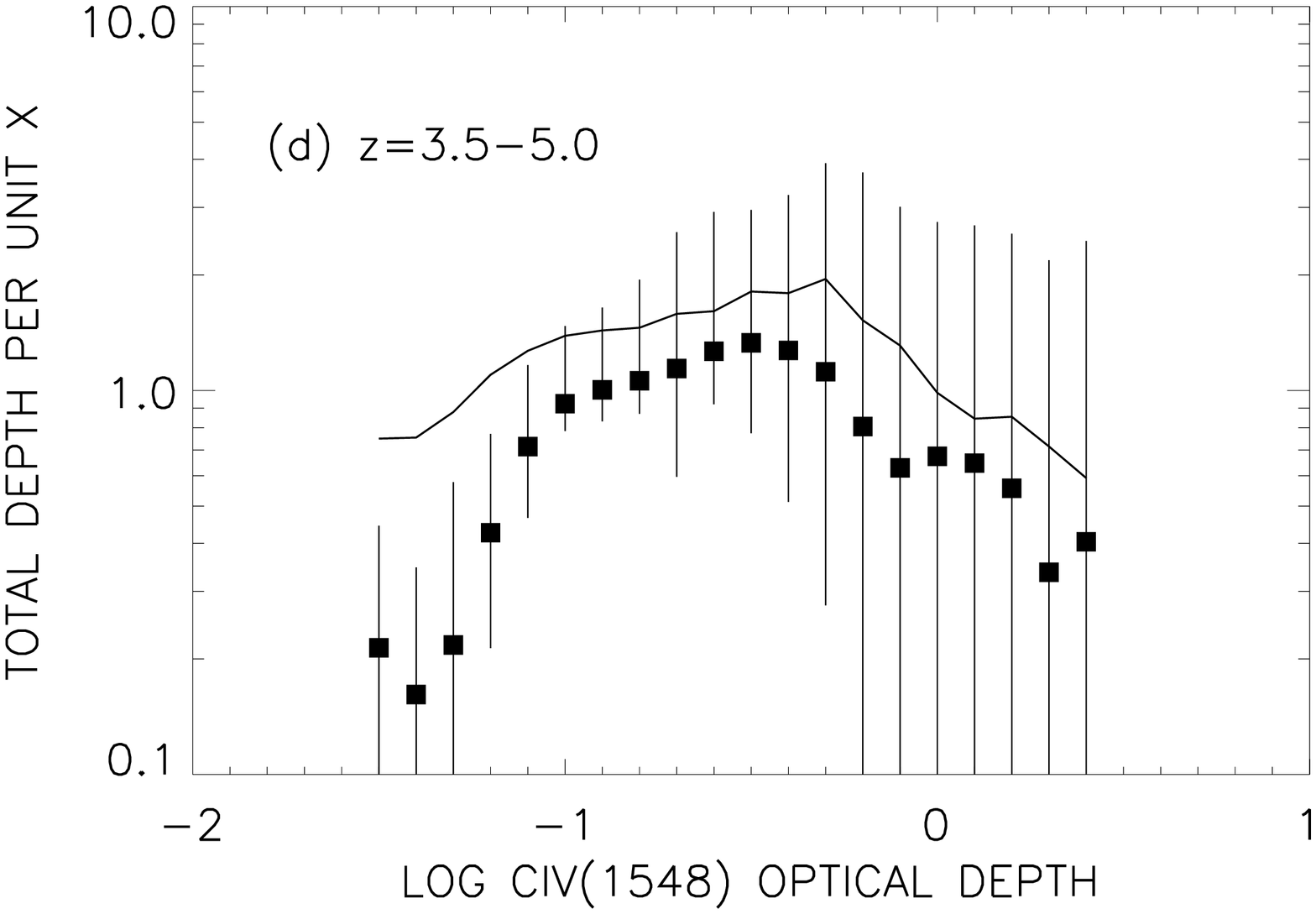}
\vspace{6pt}
\figurenum{18d}
\caption{
}
\label{fig18d}
\addtolength{\baselineskip}{10pt}
\end{inlinefigure}

%
% Figure 18e CIV o.d. distributions
%
\begin{inlinefigure}
\includegraphics[angle=90,scale=.6]{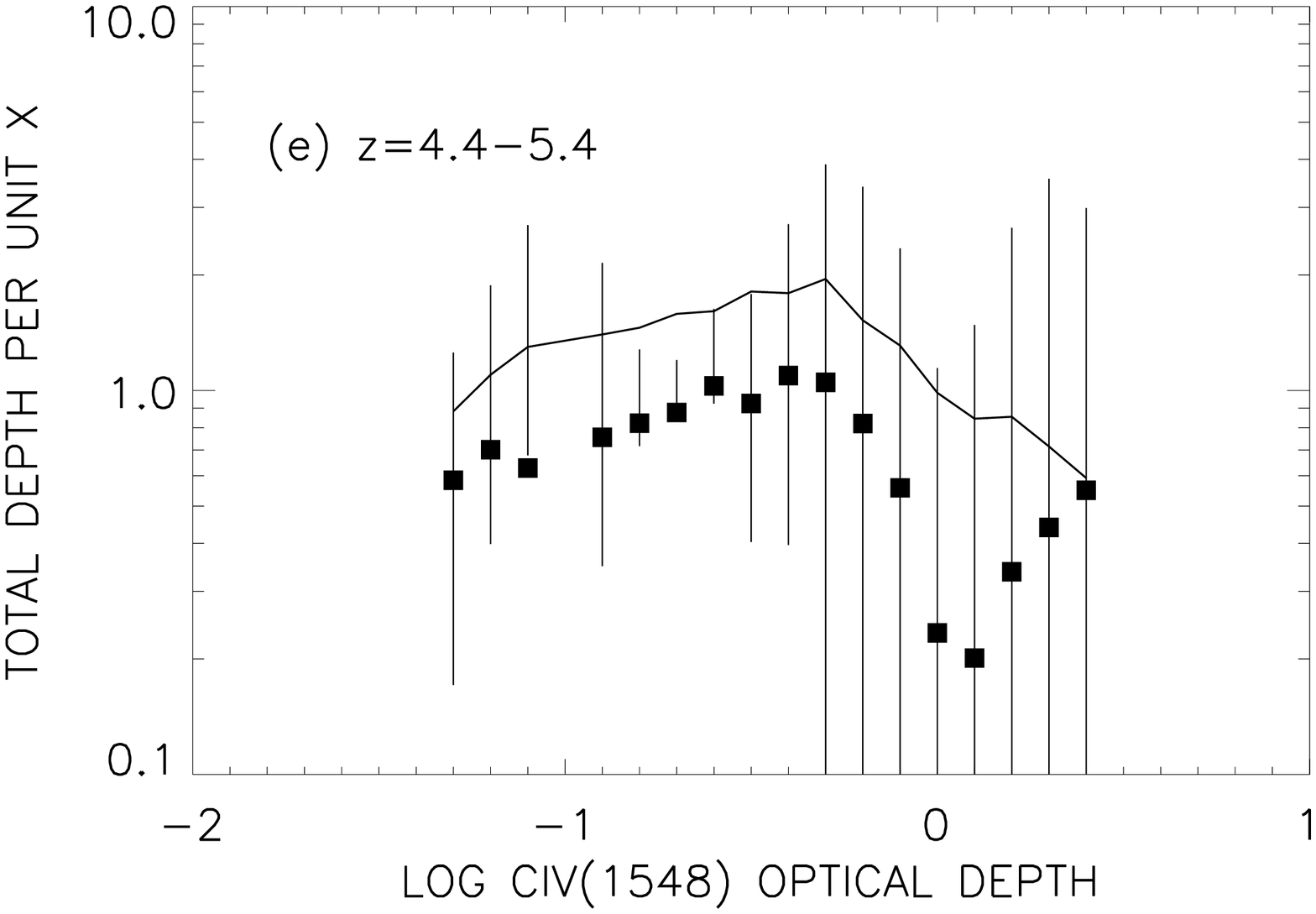}
\vspace{6pt}
\figurenum{18e}
\caption{
}
\label{fig18e}
\addtolength{\baselineskip}{10pt}
\end{inlinefigure}

\end{document}